\documentclass[a4paper,11pt]{article}
\pdfoutput=1 

\usepackage{jheppub} 

\makeatletter
\def\@fpheader{\relax}
\makeatother

\usepackage[T1]{fontenc} 

\usepackage[all,matrix,arrow,curve]{xy}
\usepackage{amssymb}
\usepackage{stmaryrd}

\usepackage{shuffle}
\usepackage[all]{xy}
\usepackage[vcentermath]{youngtab}

\def\owedge{\varowedge}
\def\be{\begin{equation}}
\def\ee{\end{equation}}
\def\ba{\begin{eqnarray}}
\def\ea{\end{eqnarray}}

\def\rd{\mathrm{d}}

\def\g{\mathfrak{g}}

\def\h{\mathfrak{h}}
\def\CN{{\cal N}}
\def\CM{{\cal M}}
\newcommand{\md}{\mathrm{d}}

\def\V{\mathbb{V}}
\def\W{\mathbb{W}}
\def\S{\mathbb{S}}
\def\gL{\g_{\mathrm{Lie}}}
\def\tL{\theta_{\mathrm{Lie}}}
\def\tto{\twoheadrightarrow}

\newcommand*{\Z}{{\mathbb Z}}
\newcommand*{\R}{{\mathbb R}}

\newcommand {\subplus}{\mathop{{\subset}\llap{\raise
0.15pt\hbox{\normalfont\small+}\hskip 0.5pt}}}

\newcommand {\subtimes}{\mathop{{\subset}\llap{\raise
0.2pt\hbox{\normalfont$\times$}\hskip -0.2pt}}}

\title{\boldmath The Embedding Tensor, Leibniz-Loday Algebras, and Their Higher Gauge Theories}

\author[a]{Alexei Kotov}
\author[b]{Thomas Strobl}

\affiliation[a]{Faculty of Science, University of Hradec Kralove, \\ Rokitanskeho 62, Hradec Kralove
50003, Czech Republic.}
\affiliation[b]{Institut Camille Jordan,
Universit\'e Claude Bernard Lyon 1 \\
43 boulevard du 11 novembre 1918, 69622 Villeurbanne cedex,
France.}

\emailAdd{oleksii.kotovATuhk.cz}
\emailAdd{stroblATmath.univ-lyon1.fr}

\abstract{We show that the data needed for the method of the embedding tensor employed in gauging supergravity theories are precisely those of a Leibniz algebra (with one of its induced quotient Lie algebras embedded into a rigid symmetry Lie algebra that provides an additional ``representation constraint''). Every Leibniz algebra gives rise to a Lie n-algebra in a canonical way (for every $n\in\mathbb{N}\cup \{ \infty \}$). It is the gauging of this $L_\infty$-algebra that
explains the tensor hierarchy of the bosonic sector of gauged supergravity theories. The tower of p-from gauge fields corresponds to Lyndon words of the universal enveloping algebra of the free Lie algebra of an odd vector space in this construction. Truncation to some $n$ yields the reduced field content needed in a concrete spacetime dimension.

\vspace*{1.0em}
}

\begin{document}
\maketitle

\section{Introduction}
The (method of the) ''embedding tensor'' and the associated ''tensor hierarchies'' is an elegant and useful way for the construction of supergravity theories and, by freezing the gravity sector, of higher gauge theories. There exists a  big literature on the subject, see, e.g., \cite{H1,H2,H3,H4,H5,H6,H7,H8,H9}.

The intention of the present note is two-fold: First, we show that every embedding tensor is canonically associated to a Leibniz algebra. A Leibniz algebra is a generalization of a Lie algebra: the bilinear product defined on a vector space $\V$, which we denote by $\circ$ in this note, is not necessarily antisymmetric, while it still satisfies an appropriate version of the Jacobi identity. In particular, if $\circ$ is antisymmetric, the axioms reduce to that of a Lie algebra. This mathematical notion was studied by Jean-Louis Loday, who showed among others that the tensor algebra ${\cal T}^\bullet \V^*$ receives the structure of a complex when equipped with an appropriate degree plus one operator named after him. What we want to stress as a first point in this article is, in particular,  that one should shift the focus from embedding tensors directly to Leibniz algebras (or Loday algebras, in honor of Jean-Louis) in the  construction of the corresponding gauge theories.

Second, we show that every Leibniz algebra gives canonically rise to  a higher homotopy version of a Lie algebra: a Lie $\infty$-algebra \cite{Lada1,Lada2,Severa}. In fact, it is known that, at least in the absence of scalar fields, every higher gauge theory underlies a structural Lie $\infty$-algebra, cf, e.g., \cite{Gruetzmann,Qbundles,Lie4}. It is thus comforting to know that there is such a canonical Lie $\infty$-algebra associated to every Leibniz algebra. The most elegant description of such an algebra is provided by what is called a differential graded (dg-) manifold $({\cal M},Q)$, called also simply a $Q$-manifold sometimes. For $n=1$, this reproduces the Chevalley-Eilenberg complex $(\Lambda \g^*, \rd_{CE},\wedge)$ for some ordinary Lie algebra  $\g$. In the present article we show that the above mentioned Loday complex ${\cal T}^\bullet \V^*$ together with the Loday differential $\mathrm{d}_L$ and an odd version of the shuffle product also define a Q-manifold. We note in parenthesis that even if the Leibniz algebra is a Lie algebra, this Q-manifold is bigger than the one corresponding to the Chevalley-Eilenberg complex. As a graded manifold, it can be identified with the free graded Lie algebra of the vector space $\V$ when shifted into degree -1. The Loday differential $\mathrm{d}_L$ induces an odd, nilpotent vector field $Q_L$ on this manifold and thus defines an infinitely extended homotopy Lie algebra. We define truncations or, more precisely, projections to finite $n$ Q-manifolds or Lie $n$-algebras. These are those needed concretely for the construction of higher gauge theories in a particular spacetime dimension. We pay particular attention to the case $n=2$, which corresponds to non-abelian gerbes (in arbitrary spacetime dimensions), displaying its 1-, 2-, and 3-brackets explicitly, and demonstrate how they are obtained by the general truncation process.

More mathematical details will be presented elsewhere. For example, applying the in part elaborate techniques of \cite{Fiorenza,Getzler} to a differential graded Lie algebra (dgla) that one can associate to every Leibniz algebra---where one of the options for this dgla is to use the Kantor algebra appearing in \cite{Palmkvist} or, simpler, a suitable sub-dgla of it---one also obtains a Lie  $\infty$-algebra. In \cite{KSmath} we will prove the non-trivial fact that it agrees with the one in our construction mentioned above.
 Also action functionals and their gauge invariance as obtained when using the findings of the present paper will be presented on another occasion---but see also \cite{gerbe,Wagemann,LeibYM} for some related steps in the case $n=2$ as well as  the last subsection to this article, section \ref{sec:new}.

\subsection{About the first version of this paper from 2013}
The main findings of this paper were obtained in 2012/2013. A first, partially finished draft  existed since 2013. It consisted of the present front page (title and abstract), section  \ref{sec:Embedding}, the subsections \ref{sec:Lodaycomplex}, \ref{sec:extension}, and Appendix \ref{sec:app} (up to small polishing, without any change of contents, and the later addition of footnotes). The content of the whole paper, except for subsection \ref{sec:new}, was moreover communicated by one of us in \cite{Luxtalk}. We originally intended to include more mathematical details about the Lie infinity structure as well as an extended part on the general construction of the associated higher gauge theories, using the framework of \cite{Gruetzmann,Qbundles}. However, the 2013 draft was available to some people and its results have been used in \cite{Lavau,Olaf-Henning}. In addition, the observation of the canonically associated Lie 2-algebra, a special case of our findings, has been meanwhile found independently in \cite{ChinesenLie2} and a related one also in \cite{Hohm}. We thus decided to round up the original text by the addition of three subsections and the above short introduction and to postpone further analysis to later work.

\newpage

\section{Embedding tensor versus Leibniz algebra \label{sec:Embedding}}
\subsection{The embedding tensor} \label{sec:theta}
In this subsection we provide a basis-independent description of the embedding tensor.

Let $(\g, [ \cdot , \cdot])$ be a Lie algebra and consider a vector space $\mathbb{V}$ carrying a representation of $\g$. An embedding tensor is a map $\theta \colon \mathbb{V} \to \g$ onto a Lie subalgebra $\g_0 \subset \g$ that is $\g_0$-equivariant. Equivariance means that for any $\xi \in \g_0$ and $v \in \V$ one has
\be
\theta(\xi \cdot v) = \xi \cdot \theta(v) \equiv [\xi , \theta(v)], \label{eq:inter}\ee i.e.~$\theta$ is an intertwiner between the $\g_0$-representations $\V$ and $\g_0$. Here we used the fact that any $\g$-representation induces a $\g_0$-representation and that any Lie algebra is a representation of itself with respect to the left-multiplication (regular representation), i.e.~with respect to the adjoint action on itself (adjoint representation).

$\theta$ can be regarded also as a $\g_0$-invariant element in the $\g$-representation $\V^* \otimes \g$, 
$\theta \in (\V^* \otimes \g)^{\g_0}$. This representation will be reducible in general and supersymmetry restricts $\theta$ to particular, invariant subspaces, which is called the \emph{representation constraint} on the embedding tensor. 
This constraint will not play an important role in the present context, however, since the considerations of the present paper are valid for any $\theta$ as defined above. In fact, all of $\g$ does not enter here, except for that $\mathrm{im} (\theta) = \g_0$ is naturally embedded into $\g$ and thus probably gave rise to the name ``embedding tensor''. Restricting the whole discussion directly to $\g_0$ for this reason, we arrive at the following simplified definition, which we will use from now on: An \emph{embedding tensor} \be \theta \colon \V  \twoheadrightarrow
\g_0  \label{theta}\ee  is a surjective intertwiner from a $\g_0$-representation $\V$ to its adjoint representation. .

The so-called \emph{closure constraint}, viewed as a quadratic equation on $\theta$, is automatically contained in this definition. To see this, we  choose a basis $e_M$ in $\V$ and $b_\alpha$ in $\g_0$, $[b_\alpha,b_\beta]=f_{\alpha \beta}{}^\gamma \, b_\gamma$; then the embedding tensor corresponds to the matrix $\theta_M{}^\alpha$ appearing in $\theta(e_M) = \theta_M{}^\alpha b_\alpha$.  Furthermore, denote the representation matrices corresponding to $b_\alpha$ by $t_\alpha \in \mathrm{End}{\V}\cong \mathrm{Mat}_{\dim(\V)}$, so that $b_\alpha \cdot e_M = -t_{\alpha M}{}^N e_N$. ($b_\alpha \cdot$ denotes the $b_\alpha$ in the representation, the minus sign in the last equation ensures the same for the matrices $t_\alpha$, i.e.~the matrix commutator satisfies $[t_\alpha,t_\beta]=f_{\alpha\beta}{}^\gamma t_\gamma$).\footnote{For a simpler comparison with the literature, we adapt the notation correspondingly, cf., e.g., \cite{H7}.} 
Then the condition (\ref{eq:inter}) becomes $\xi^\alpha v^M \theta(t_{\alpha M}{}^N e_N) = \xi^\alpha v^M [b_\alpha , \theta_M{}^\beta b_\beta]$. Making use of the fact that $\theta \colon \V \to \g_0$ is surjective, we can replace $\xi^\alpha$ by $\theta_P{}^\alpha w^P$. The resulting equation has to hold for all $v,w \in \V$ and thus indeed reduces to the known closure constraint
\begin{equation}
\theta_P{}^\alpha t_{\alpha M}{}^N \theta_N{}^\gamma +\theta_P{}^\alpha \theta_M{}^\beta f_{\alpha \beta}{}^\gamma =0\, .
\end{equation}

In what follows, we will show that the embedding data as defined above can be described concisely by a more straightforward notion for the present purposes, namely the one of a Leibniz algebra on $\V$.

\subsection{Reformulation as a Leibniz algebra}
A Leibniz or Leibniz-Loday algebra $(\V, \circ)$ is a vector space $\V$ together with a bilinear product $\circ$ satisfying for all $x,y,z \in \V$
\begin{equation} \label{eq:Leib}
x \circ (y \circ z) = (x \circ y )\circ z+ y \circ (x \circ z) \, .
\end{equation}
Thus the left-multiplication with respect to any element is a derivation of the product. If $\circ$ is in addition antisymmetric, the Leibniz algebra becomes a Lie algebra; so any Lie algebra is a Leibniz algebra, but certainly not vice versa.

We now collect some simple facts about Leibniz algebras and show how each Leibniz algebra $(\V,\circ)$ induces an embedding tensor. Thereafter we will show that also vice versa any embedding tensor gives rise to a Leibniz algebra, concluding with remarks under which conditions the composition of the two directions reproduces the respective starting point.

Let us split the product $m \colon \V \otimes \V \to \V$, $x \otimes y \mapsto x\circ y$, into its symmetric and antisymmetric part:
 \begin{eqnarray}
  s \colon S^2 \V \to \V &,& x \vee y \mapsto  x\circ y + y\circ x=: 2\left(x \bullet y\right)
  \label{s}\\
  a \colon \Lambda^2 \V \to \V &,& x \wedge y \mapsto x\circ y - y\circ x  =: 2\left(x \star y\right)\label{a}
 \end{eqnarray}
so that $x \circ y =  x \bullet y+ x \star y$.\footnote{We use conventions where $ x \wedge y = x \otimes y - y \otimes x$ and $ x \vee y = x \otimes y + y \otimes x$.} Let  $\S:= s(S^2 \V)\subset \V$ denote the vector subspace of Leibniz squares. It is easy to see that $\S$ is a both-sided ideal in $\V$; e.g.~setting $y=x$ in (\ref{eq:Leib}), one finds
\begin{equation} \label{eq:bullet}
(x \bullet x) \circ z = 0    \qquad \forall x,z \in \V \, .
\end{equation}
Quotienting $\V$ by the ideal $\mathrm{im}\, s = \S$ provides a vector space $\gL\equiv \V/\S$ on which the induced product is antisymmetric and thus Lie. Denoting by $\tL \colon \V \twoheadrightarrow \gL$, $x \mapsto [x]$ the corresponding quotient map and the induced product by a Lie bracket, $[x \circ y] := [ [x] , [y] ]$,  we thus arrive at the following exact sequence
\begin{equation}\label{eq:exact}
0 \to \W_{\mathrm{Lie}} \stackrel{t_{\mathrm{Lie}}}{\longrightarrow} (\V , \circ ) \stackrel{\tL}{\longrightarrow} (\gL , [\cdot , \cdot ]) \to 0 \, ,
\end{equation}
where  $\W_{\mathrm{Lie}} := S^2 \V / \ker s$ and $t_{\mathrm{Lie}}$ denotes the natural isomorphism of this vector space with its image $\S$ inside $\V$. Since the restriction of the Leibniz product $\circ$ to two elements of $\S$ vanishes according to \eqref{eq:bullet}, it is natural to equip $\W_{\mathrm{Lie}}$ with the structure of an abelian Lie algebra. In this way, \eqref{eq:exact} becomes an exact sequence of Leibniz algebras, in which $\W_{\mathrm{Lie}}$ and $\gL$ are even Lie algebras.

Eq.~(\ref{eq:bullet}) has another consequence: there is a natural action of $\gL$ on $\V$. Indeed, the ambiguity of a preimage of $\tL(x)$ lies inside the image of $s$ or $t_{\mathrm{Lie}}$ and thus acts trivially on an element of $\V$ from the left according to this equation. The rest then follows from the Leibniz property (\ref{eq:Leib}): $(x + \S) \circ \left((y + \S) \circ z \right) - (x \leftrightarrow y)  = (x \circ y) \circ z = (x \circ y + \S) \circ z$, so that indeed $[ \tL(x) \cdot , \tL(y) \cdot ] z$, the commutator of the action of two elements in $\gL = \mathrm{im} \, \tL$ on $z\in\V$, is equal to  $\tL(x \circ y) \cdot z = ([\tL(x),\tL(y)]) \cdot z$, the action of the Lie bracket of these two elements on $z$.

The notation is already close to the one used in the previous subsection: Just rename $\gL$ to $\g_0$ and $\tL$ to simply $\theta$. It then only remains to show equivariance (\ref{eq:inter}) of the map $\theta$, i.e.~$\theta(\theta(x) \cdot y) = [\theta(x),\theta(y)]$ for all $x,y\in\V$. This, however, is nothing but the homomorphism property of the quotient map, $\theta((x+\S) \circ y) = \theta(x \circ y) = [\theta(x),\theta(y)]$. Thus the map $\tL \colon \V \twoheadrightarrow \gL$ or, in the simplified notation, $\theta \colon \V \twoheadrightarrow \g_0$, together with the $\g_0$-action on $\V$ defines an embedding tensor.

So, each Leibniz algebra $(\V,\circ)$ is seen to induce an embedding tensor in a canonical way. We will call the embedding tensor $\tL\colon \V \twoheadrightarrow \gL$ universal, for reasons to become clear shortly.

The reverse direction is even simpler: Given an embedding tensor $\theta \colon \V \twoheadrightarrow \g_0$ as defined in the previous subsection, then $\V$ can be equipped canonically with a product $\circ$ satisfying equation (\ref{eq:Leib}). Indeed, one may simply define $\forall x,y \in \V$
\begin{equation} \label{eq:reconstr} x \circ y := \theta(x) \cdot y \, .
\end{equation}
The Leibniz property now follows from the representation and equivariance conditions: $x\circ (y\circ z) - (x\leftrightarrow y) = \left([\theta(x),\theta(y)]\right) \cdot z = \theta(\theta(x) \cdot y)\cdot z=(x\circ y)\circ z$.

We now address the question of equivalence of the two notions. If one starts with a Leibniz algebra $(\V,\circ)$, determines its embedding tensor $\tL$ in  (\ref{eq:exact}), and reconstructs from the resulting data the Leibniz product by means of (\ref{eq:reconstr}), one indeed returns to the original Leibniz algebra. Performed in this order, the two procedures are inverse to one another.

If, on the other hand, we start with an embedding tensor $\theta \colon \V \twoheadrightarrow \g_0$, it determines its \emph{unique} Leibniz structure $(\V,\circ)$ according to (\ref{eq:reconstr}), which in turn determines the canonical embedding tensor $\tL \colon \V \twoheadrightarrow \gL$. In general, one does not obtain equality with the original embedding tensor in this way. One always has $\S \subset \ker\theta \subset \V$, but it may happen that $\S$  is strictly smaller than $\ker\theta$. This shows, however, that always $\S$ is embedded into $\W:=\ker\theta$.
The embedding tensor or intertwiner $\tL$ is universal in the usual (mathematical) sense: If $\theta \colon  \V \twoheadrightarrow \g_0$ is an embedding tensor for the Leibniz algebra $(\V,\circ)$, then this intertwining map always factors through the canonical intertwiner $\tL \colon  \V \twoheadrightarrow \gL$, $\theta = \mathrm{pr} \circ \tL$, where $ \mathrm{pr}\colon \gL \twoheadrightarrow \g_0$ is the (natural and equivariant) projection to the quotient of $\gL$ by an ideal. The situation is illustrated in the following diagram:
\vskip -2mm
\hskip3cm
\xymatrix{ 0\ar[dr]\\
& \W \ar[dr] \\ 0\ar[r] & \S \ar@{^{(}->}[u] \ar[r] & \V \ar[dr]^\theta \ar[r]^{\tL} &
 \mathfrak{g}_{\scriptscriptstyle Lie} \ar[d]^{\mathrm{pr}} \ar[r] & 0
\\ &&& \mathfrak{g}_0\ar[d] \ar[dr] \\ &&& 0 & 0}
\vskip 2mm
\emph{Figure 1: For a given Leibniz algebra $\V$, every ''embedding tensor'' (intertwiner) $\theta$ giving rise to it factors through the universal embedding tensor $\tL$.}
\vskip 2mm
\newpage

The smallest possible $\g_0$ is the one that acts effectively on $\V$. So, if we want to obtain a strict equivalence between embedding tensors as defined in the previous subsection and Leibniz algebras $(\V,\circ)$, we have to require that the $\g_0$-action on $\V$ is effective. Note that, at least in the absence of an embedding Lie algebra $\g\supset \g_0$, which anyway does not play a role in the present considerations, this may be also a good working definition for a unique ``minimal'' embedding tensor: If $\g_0$ does not act effectively on $\V$, i.e.~the representation map $\rho \colon \g_0 \to \mathrm{End}(\V)$ has a non-trivial kernel, its elements $\ker \rho$ form an ideal ${\cal I} \subset \g_0$. Quotienting by this ideal yields another embedding tensor, $\theta_{\mathrm{min}} \colon \V \twoheadrightarrow \g_{\mathrm{min}}=\g_0/{\cal I}$. For a given Leibniz algebra, any embedding tensor $\theta \colon \V \twoheadrightarrow \g_0$ lies between the universal and the minimal one, i.e.~$ \gL \to \g_0\to \g_{\mathrm{min}}$ (and correspondingly for the intertwiners, as is easy to induce from the above diagram).\footnote{There is a one-to-one correspondence between embedding tensors as defined in this paper and what is termed a Leibniz couple $\mathfrak{i} \subset \V$ in \cite{Wagemann}. Here $\mathfrak{i}$ is an ideal in  $(\V,\circ)$ fitting into the sequence of vector spaces $\S \subset \mathfrak{i} \subset {\cal Z}_L(\V)$, where ${\cal Z}_L(\V)$ denotes the left-center of $\V$. \label{newfootnote}In particular, the universal and the minimal  embedding tensor of a given Leibniz algebra correspond to  $\V/\S =  \gL$ and  $\V/ Z_L(\V) = \g_{\mathrm{min}}$, respectively.}

We illustrate the situation with some simple examples: Take a Lie algebra $\g_0$ having a non-trivial center ${\cal Z}[\g_0]\subset \g_0$. Set $\V = \g_0$ and take as an embedding tensor the identity map.
The Leibniz algebra $(\V,\circ)$ in this example is just the original Lie algebra $(\g_0,[\cdot , \cdot])$. The action of $\g_0$ on $\V$ is non-effective precisely due to the existence of a center ${\cal Z}[\g_0]$. Take the quotient $\g_1 := \g_0 / {\cal I}_1$ where ${\cal I}_1 \subset
{\cal Z}[\g_0]$ and let $\theta_1 \colon \V \equiv \g_0 \to \g_1$ be the canonical projection map, that is another embedding tensor. The ``minimal'' embedding tensor of this example arises if we quotient out all of $\ker \rho \equiv {\cal I} = {\cal Z}[\g_0]$, $\theta_{\mathrm{min}} \colon \V \twoheadrightarrow \g_{\mathrm{min}}=\g_0/{\cal I}$. Although it can happen that $\g_1$ again has a center ${\cal Z}[\g_1]$, we now can no more repeat the quotient procedure because its elements act necessarily non-trivially on $\V=\g_0$ (think, e.g., of the Heisenberg Lie-algebra $\h$, where its quotient $\h_{\mathrm{min}}=\h/{\cal Z}[\h]$ is abelian and thus equal to its own center ${\cal Z}[\h{\mathrm{min}}]$, but it acts effectively on the original $\V=\h$ by translations).

We conclude this set of examples by remarking that certainly one can quotient Lie algebras $\g_0$ by ideals that act effectively on $\V$, still obtaining an equivariant surjective map: Take e.g.~$\V = \g_0 := \g_1 \oplus  \g_2$ where $\g_1$ and $\g_2$ are e.g.~both isomorphic to $\mathrm{sl}_2$ and let $\g_0$ act again on $\V$ by adjoint transformations. In this example, $\ker \rho = 0$ and $\g_0 = \g_{\mathrm{min}}$. Denote the corresponding embedding tensor or equivariant surjective intertwiner by $\theta_{min} \colon \V \to \g_0$. Still, we can consistently quotient $\g_0$ by the ideal $\g_2$, obtaining another equivariant map and thus an apparently even ``more minimal'' embedding tensor $\theta \colon \V \tto \g_1$. However, in this example, the two embedding tensors correspond to different Leibniz algebras, which are both Lie algebras here. In the first case, $(\V,\circ) \cong \g_0= \g_1 \oplus  \g_2$, and in the second case $(\V,\circ) \cong \g_1 \oplus \R^3$, where $\R^3$ denotes the abelian three-dimensional Lie algebra. In the latter case, $\g_1$ acts trivially on the $\R^3$, moreover, and a minimal embedding tensor for \emph{this} Leibniz algebra is obtained by factoring out $\R^3$.  Since in this example $\g_1$ is a Lie sub-algebra of $\g_1 \oplus \g_2$, this illustrates also the possible role which the choice of the ``gauging'' Lie algebra $\g_0$ inside the rigid symmetry Lie algebra $\g$ can play at the very beginning.

To our mind, it is not necessary  to arrive at a bijection between embedding tensors and Leibniz algebras (although possible if one adds an additional condition as mentioned above---but cf.\ also footnote \ref{newfootnote} above). All what follows, the construction of an $L_\infty$-algebra and its corresponding higher gauge theories, only depend on the Leibniz algebra $(\V,\circ)$, which is possibly a somewhat exotic but after all a simple notion, moreover, and which is uniquely determined by a given embedding tensor (cf.~Fig.~1 above). We thus argue for a shift of attention from embedding tensors to Leibniz algebras in the context of gauged supergravity theories, and, if one wants to go further, to  $L_\infty$-algebras and Q-bundles \cite{Qbundles}---where, however, Leibniz algebras contain some additional information useful for the construction of functionals. We will come back to this in the subsequent section.

For comparison with the literature we use again a basis $\{e_M\}_{M=1}^{\dim \V}$ of $\V$. Then the Leibniz product can be characterized by structure constants,
\begin{equation}\label{eq:X}
e_M \circ e_N = X_{MN}{}^P\, e_P \, .
\end{equation}
If, as before, $T_\alpha$ denotes the representation of the generators $b_\alpha$ of $\g_0$ and $T_\alpha (e_N) = T_{\alpha N}{}^P e_P$, then from eq.~(\ref{eq:reconstr}) we obtain
\begin{equation}\label{eq:factor}
X_{MN}{}^P = \theta_M{}^\alpha T_{\alpha N}{}^P \, .
\end{equation}
From the above considerations we learn that a factorization of the structure constants of a Leibniz algebra into a product as in (\ref{eq:factor}) containing the ``embedding tensor'' is not a restriction, but always possible. And in any case, it is $(\V,\circ)$ or the structure constants  $X_{MN}{}^P$, satisfying the defining Leibniz property
\begin{equation}
X_{MK}{}^S X_{NS}{}^L - (M \leftrightarrow N)= - X_{MN}{}^S X_{SK}{}^L \, ,
\end{equation}
that we will use in all what follows.

\subsection[The associated Lie 2-algebra]{The associated Lie 2-algebra\protect\footnote{In the notation of foonote \ref{newfootnote}, there is a Lie 2-algebra for every Leibniz couple $\mathfrak{i} \subset \V$, with the extreme choices $\mathfrak{i} = \S$ and $\mathfrak{i} = {\cal Z}_L(\V)$ being canonically associated to the Leibniz algebra. In the mean time there appeared two other articles, \cite{ChinesenLie2} and \cite{Hohm} with related results: The Lie 2-algebra in \cite{ChinesenLie2} corresponds to the choice $\W \cong \mathfrak{i} := {\cal Z}_L(\V)$, and all smaller choices for $\mathfrak{i}$ sit inside this one. We will show in section \ref{sec:truncation}  that they in turn sit inside the one constructed in \cite{Hohm}. The Lie 2-algebra appearing in the (truncated) tensor hierarchy of physical interest, on the other hand, is the smallest one of all these, the one with $\W \cong \S$.\label{footnoteHohm}}}

\label{sec:Lie2}
Given a Leibniz algebra $(\V,\circ)$, the antisymmetric part $\star$ of the product does in general not provide a Lie bracket. The Jacobiator $J(x,y,z) := x \star (y\star z) + \mathrm{cycl}(x,y,z)$ might be non-zero. The violation is rather mild, however: Since $0 \to \W \stackrel{t}{\to} \V \stackrel{\theta}{\to} \g_0 \to 0$ is an exact sequence of Leibniz algebras, and on the r.h.s.~one has a Lie algebra, i.e.~a vanishing Jacobiator, the Jacobiator $J$ on $\V$ must lie in the image of the embedding map $t$. In addition, the Jacobiator as defined above is totally skew-symmetric. This implies that there must exist an $H \in \Lambda^3 \V^* \otimes \W$ such that
\begin{equation}
J(x,y,z) = t(H(x,y,z)) \, . \label{tH}
\end{equation}
All together,
\begin{equation}
\W \stackrel{t}{\to} \V  \label{VW}
\end{equation}
is equipped with what is called a Lie 2-algebra or a 2-term $L_\infty$-algebra. Let us explain this in more detail. Let us declare elements of $\V$ to have degree $-1$ and elements of $W$ degree $-2$. We now need multilinear brackets  of degree +1 on this short complex (in this convention of describing an $L_\infty$-algebra). In fact, for a 2-term $L_\infty$-algebra we need the $k$-ary operations only for $k=1,2,3$. The map $t\colon \W\to \V$ has degree +1 and can serve as the 1-bracket, extended by zero when acting on elements of degree $-1$. Note that $t$ is just an embedding and thus elements of $\W$ can be viewed of as elements in $\V$, but with the degree shifted by one. To distinguish these two from one another,  we will denote an element $x \in \mathrm{im}(t)\subset \V$ by an additional tilde if viewed as an element in $\W$:
$$ \tilde{x} = t^{-1}(x)\in \W. $$ The 2-bracket can now be defined as follows (recall the notation introduced in Eqs.\ \eqref{s} and \eqref{a} above):  for all $x,y \in \V$ and  $\tilde{x},\tilde{y}\in \W$ we put
\begin{equation} \label{coucou}
[x,y] := x \ast y  \: , \qquad [x,\tilde{y}] :=  \widetilde{x \bullet y}   \: , \qquad [\tilde{x},\tilde{y}]:=0 \, .
\end{equation}
The last equation is enforced by degree reasons, since the vector space of degree $-3$ is by definition zero.
Finally, since $H(x,y,z)$ takes values in $\W$ and elements of $\V$ have degree $-1$, also $H$ has degree +1 and can be considered to be the only non-vanishing part of the 3-bracket $[\cdot,\cdot,\cdot]$. By an explicit computation of the Jacobiator, one obtains
\begin{equation}\label{Jac'}
J(x,y,z) = -\tfrac{1}{3} x \bullet (y\ast z) + \mathrm{cycl}(x,y,z) \, .
\end{equation}
Thus, for $H$ in Eq. (\ref{tH}) or the 3-bracket of the Lie 2-algebra, we find \begin{equation}
 [x,y,z] := H(x,y,z)= -\tfrac{1}{3} \widetilde{x \bullet (y\ast z)} + \mathrm{cycl}(x,y,z) \, ,
\end{equation}
with all other components of the 3-bracket vanishing, $[\tilde{x}, \cdot, \cdot] := 0$ for all $\tilde{x} \in \W$.  We remark in parenthesis, that some authors prefer the notion $l_1,l_2$, and $l_3$ for $t$, $[\cdot, \cdot]$, and $[\cdot, \cdot,\cdot]$, respectively, calling $l_1$ the 1-bracket of the $L_\infty$ algebra.

It is a non-trivial and somewhat lengthy calculation to verify that the above data indeed satisfy all the higher Jacobi identities needed for the definition of a Lie 2-algebra, which is a 2-term $L_\infty$ algebra. There will be, however, a much more elegant way of establishing this result without the need of a direct verification for the case $\W \cong \S$. In the subsequent section we will present a Lie infinity algebra canonically associated to every Leibniz algebra. Its consistent truncation to a Lie 2-algebra will be seen to coincide with this Lie 2-algebra then.

\newpage
\section{Associated Lie infinity algebras and their gauge theories}\noindent

\subsection{The Loday complex}
 \label{sec:Lodaycomplex}
 Any Lie algebra $(\g,[\cdot,\cdot])$ gives rise to the Chevalley-Eilenberg complex $(\Lambda^\bullet \g^*, \rd_{CE})$. For $\omega \in \Lambda^p \g^*$ it is defined by means of
\begin{equation}\label{eq:dCE}
\left(\rd_{CE}\, \omega\right)(x_1,\ldots,x_{p+1})=\sum_{1\leq i<j\leq p+1}(-1)^{i+j}\omega([x_i,x_j], x_1, \ldots , \widehat{x_i}, \ldots, \widehat{x_j}, \ldots, x_{p+1})\,,
\end{equation}
where the over-hat signals omission and $x_1, \ldots, x_{p+1} \in \g$. $\rd_{CE}$ squares to zero, if and only if the product $[\cdot, \cdot]$ satisfies the Jacobi identity $J(x_1,x_2,x_3)=0$. In particular, it does not for a Leibniz algebra; in fact, for a general Leibniz algebra $(\V,\circ)$, there are two obvious ways of interpreting a formula of the type (\ref{eq:dCE}): one, where we replace the Lie bracket by the antisymmetric part of the Leibniz product. Then, however, the non-vanishing Jacobiator obstructs the operator to square to zero. On the other hand, if we replace the Lie bracket by the Leibniz product $\circ$, the operator does not map antisymmetric tensors into antisymmetric tensors in general.

Loday observed \cite{Loday1,Loday2} that every Leibniz algebra $(\V,\circ)$ gives rise to a complex by using the whole tensor algebra ${\cal T}^\bullet \V^* = \V^* \oplus \V^* \otimes \V^* \oplus \ldots$ together with the differential\footnote{Loday considered right-Leibniz algebras $(\V, \bar{\circ} )$ where by definition $(x \,\bar{\circ}\, y) \, \bar{\circ} \,z =  x\, \bar{\circ} \,(y \, \bar{\circ} \,z) + (x\,\bar{\circ} \,z) \,\bar{\circ}\, y$, which then requires a slightly different definition of the differential. In general, a right-Leibniz algebra is not simultaneously also a Leibniz algebra $(\V, \circ)$ in our sense, i.e.~a left-Leibniz algebra. However, the relation $x \circ y := y \,\bar{\circ} \,x$ provides a natural bijection between these two notions.}
\begin{equation}\label{eq:dL}
\left(\rd_{L}\, \omega\right)(x_1,\ldots,x_{p+1})=\sum_{1\leq i<j\leq p+1}(-1)^{i+1}\omega(x_1, \ldots , \widehat{x_i}, \ldots,x_i\circ x_j, \ldots, x_{p+1})\,,
\end{equation}
for any $\omega \in {\cal T}^p \V^* \equiv (\V^*)^{\otimes p}$ and $x_1, \ldots, x_{p+1} \in \V$; here the entry $x_i \circ x_j$ is at the $(j-1)^\mathrm{st}$ slot of the multilinear form $\omega$ inside the double sum. For a 1-tensor $\omega \in \V^*$, one has, in particular, $\rd_L \omega(x_1,x_2)= \omega(x_1 \circ x_2)$, while on a 2-tensor the above formula yields $\rd_L \omega(x_1,x_2,x_3)=\omega(x_1 \circ x_2,x_3) + \omega(x_2,x_1\circ x_3) - \omega(x_1,x_2\circ x_3)$. One verifies that the Loday- (or Leibniz-) operator $\rd_L$ squares to zero, if and only if the product $\circ$ satisfies the Leibniz identity (\ref{eq:Leib}); the necessity of   (\ref{eq:Leib}) becomes clear already from the previous two formulas for 1- and 2-tensors.

It is in some cases useful to consider also the dual  of the coboundary operator $\rd_L$, the boundary operator $\partial_L:=\rd_L^*\colon  {\cal T}^p \V \to {\cal T}^{p-1} \V$. By definition, for every $\omega \in  {\cal T}^p \V^*$ and every $\phi \in  {\cal T}^{p+1} \V$, one has $\langle \rd_L \omega, \phi \rangle = \langle \omega, \partial_L \phi \rangle$. Evidently, on inspection of the previous formulas,  
$\partial_L (x_1 \otimes x_2)=x_1 \circ x_2$ and $\partial_L (x_1 \otimes x_2 \otimes x_3)=(x_1 \circ x_2) \otimes x_3 + x_2 \otimes (x_1\circ x_3) - x_1\otimes (x_2\circ x_3)$.

 Denoting by
$(e^M)$ a basis dual to $(e_M)$, one has $\rd_L e^M = X_{NP}{}^M e^N \otimes e^P$, where the structure constants were introduced in (\ref{eq:X}). This reminds of a similar formula for differential forms and the de Rham differential: If the commutator of a basis of vector fields $(e_a)$ satisfies $[e_a,e_b]=C^c_{ab} e_c$ for some structure functions $C^c_{ab}$, then the for the dual basis $(e^a)$ one has, $\rd e^a = C^a_{bc} e^b \wedge e^c$. However, while the de Rham differential is extended to higher form degrees by a (graded) Leibniz rule, e.g., $\rd (e^a \wedge e^b) = \rd e^a \wedge e^b - e^a \wedge \rd e^b$, this is not the case for $\rd_L$ with respect to $\otimes$. For example,
\begin{eqnarray}
\rd_L (e^I \otimes e^J) &=&  X_{PQ}{}^Ie^P \otimes e^Q \otimes e^J+X_{PQ}{}^Ie^P \otimes e^J \otimes e^Q - X_{PQ}{}^J  \, e^I \otimes e^P \otimes e^Q\nonumber \\
&\equiv&\left( X_{KM}{}^I \delta_N^J + X_{KN}{}^I \delta_M^J  - X_{MN}{}^J \delta_K^I\right) \, e^K \otimes e^M\otimes e^N\, ,\label{dLIJ}
\end{eqnarray}
and, more generally,
\begin{equation}
\rd_L ( e^{M_1} \otimes \ldots \otimes e^{M_p})=\!\!\!\!\!\sum_{1\leq i<j\leq p+1}\!\!\!\!(-1)^{i+1} X_{P_iP_j}{}^{M_i}\: e^{M_1} \otimes \ldots  \otimes e^{P_i} 
\otimes \ldots  
 \otimes e^{P_j} \otimes e^{M_{j}}\otimes
\ldots \otimes e^{M_p}\, ,
\end{equation}
where the insertion $e^{P_i}$ replaces $e^{M_i}$ at the i-th position, $e^{P_j}$ appears at the j-th position, followed directly by $e^{M_j}$ (at the (j+1)st position), which is dropped when the index $j$ takes the value $p+1$.


\subsection[Extension to a differential graded algebra]{Extension to a differential graded algebra\protect\footnote{As a reaction to the presentation \cite{Luxtalk} of our results, we learned that such a dga was constructed recently also in \cite{Poncin}. Some of the statements in this and the subsequent subsection will be given without an explicit proof, for which we refer to \cite{KSmath}---but see also Appendix \ref{sec:app}.}} \label{sec:extension}

The Chevalley-Eilenberg operator $\rd_{CE}$ is simultaneously a differential with respect to the wedge product, i.e.~for $\omega_1 \in  \Lambda^p \g^*$ and $\omega_2 \in  \Lambda^\bullet \g^*$ one has
\begin{equation}
 \rd_{CE} (\omega_1 \wedge \omega_2) =  \rd_{CE} \,\omega_1 \wedge \omega_2 + (-1)^p \omega_1 \wedge  \rd_{CE}\,\omega_2 \, .
\end{equation}
In contrast, the operator $\rd_L$ does not have a likewise property with respect to the natural tensor product $\otimes$. This is not so surprising, since also $\rd_{CE}$ does not have this property when we use the ordinary tensor product on the alternating forms $\Lambda^\bullet \g^*\subset  {\cal T}^\bullet \g^*$.

It is surprising that Loday's celebrated observation that $({\cal T}^\bullet \V^*,\rd_L)$ defines a complex was not extended to a differential graded commutative, associative algebra structure $({\cal T}^\bullet \V^*,\rd_L,\varowedge)$  (dga for short). To find a such a compatible product $\varowedge$ on the tensor algebra, we first observe that the operator $\rd_L$ of eq.~(\ref{eq:dL}) agrees with the Chevalley-Eilenberg differential $\rd_{CE}$, recalled in eq.~(\ref{eq:dCE}), when restricted to alternating forms.
The latter one being compatible with the wedge product, it is natural to look for an extension of the wedge product $\wedge$ defined on $\Lambda^\bullet \V^*\subset{\cal T}^\bullet \V^*$ to all of the tensor algebra. Such an extension is not unique certainly. However, the following one turns out to have all the desired properties: Let $\omega_1 \in {\cal T}^p\V^*$, $\omega_2 \in {\cal T}^q\V^*$, $x_1,\dots, x_{p+q} \in \V$, and put
\begin{equation}\label{eq:owedge}
\left(\omega_1 \varowedge \omega_2\right) (x_1, \ldots, x_{p+q}) := \sum_{\sigma \in \mathrm{Sh}_{p,q}} \mathrm{sgn}(\sigma) \omega_1(x_{\sigma(1)}, \ldots , x_{\sigma(p)}) \omega_2(x_{\sigma(p+1)}, \ldots , x_{\sigma(p+q)})  .
\end{equation}
Here $\mathrm{Sh}_{p,q}=\{\sigma \in S_{p+q} | \sigma(1) < \sigma(2) < \ldots < \sigma(p),  \sigma(p+1) < \sigma(p+1) <\ldots < \sigma(p+q)\}$ are the $(p,q)$-shuffles, a subgroup of the symmetric group or permutations of $p+q$ elements, and the sign $\mathrm{sgn}(\sigma)$ is just its respective parity as permutation. This formula implies, e.g., for $\omega_1 \in \V^*$, $\omega_2 \in \V^* \otimes \V^*$, and $x_1,x_2,x_3 \in \V$,
\begin{equation} \left(\omega_1 \varowedge \omega_2\right) (x_1, x_2, x_3) = \omega_1(x_1) \omega_2(x_2,x_3) - \omega_1(x_2) \omega_2(x_1,x_3) + \omega_1(x_3) \omega_2(x_1,x_2)\, .\label{12}
\end{equation}

It is easy to verify that this product is super-commutative
\begin{equation}
\omega_1 \varowedge \omega_2 = (-1)^{pq} \omega_2 \varowedge \omega_1 \, .
\end{equation}
It  is also associative, $\omega_1 \varowedge (\omega_2 \varowedge \omega_3)=(\omega_1 \varowedge \omega_2) \varowedge \omega_3$,   which one can either verify by ``brute force'' using the definition (\ref{eq:owedge}) or by means of the more elegant method employing strings taking values in the odd vector space $\Pi \V^*$, as shown in Appendix \ref{sec:app}. Most important for our purpose is, however,  the compatibility
\begin{equation} \label{eq:compat}
\rd_{L} \left(\omega_1 \owedge \omega_2\right) = \rd_{L} \, \omega_1 \owedge \omega_2 + (-1)^p \omega_1 \owedge  \rd_{L}\, \omega_2 \, ,
\end{equation}
which turns $\rd_L$ into a differential and thus $({\cal T}^\bullet \V^*, \owedge, \rd_L)$ into a dga.

The product (\ref{eq:owedge}) is a graded version of the shuffle product. This becomes most obvious when expressed in terms of a basis: $(e^{M_1} \otimes \ldots \otimes e^{M_p}) \owedge (e^{N_1} \otimes \ldots \otimes e^{N_q})$ gives the sum of all shuffles between the elements of the two brackets, where one picks up a minus sign for each permutation of two basis elements of $\V^*$ (i.e.~one may consider $e^M \in \Pi \V^*$ for this purpose). So, the above product gives $e^{M_1} \otimes \ldots \otimes e^{M_p} \otimes e^{N_1} \otimes \ldots \otimes e^{N_q}\, - \,e^{M_1} \otimes \ldots \otimes e^{N_1} \otimes e^{M_p} \otimes \ldots \otimes e^{N_q} + \\
+e^{M_1} \otimes \ldots \otimes e^{N_1} \otimes e^{M_{p-1}} \otimes e^{M_p} \otimes \ldots \otimes e^{N_q}\,-\,e^{M_1} \otimes \ldots \otimes e^{N_1} \otimes e^{M_{p-1}}\otimes e^{N_2} \otimes e^{M_p} \otimes \ldots \otimes e^{N_q}\pm \ldots$. This gives, for example,
\begin{equation} \label{shuffle12}
e^K \owedge (e^M \otimes e^N) = e^K \otimes e^M \otimes e^N - e^M \otimes e^K \otimes e^N + e^M \otimes e^N \otimes e^K \, ,
\end{equation}
which in turn is readily verified to coincide with the previous formula \eqref{12} by setting $\omega_1= e^K$ and $\omega_2=e^M \otimes e^N$.
We will thus henceforth call $\owedge$ a \emph{graded} or \emph{odd shuffle product}.

\subsection{The associated Lie infinity algebra and the infinite tensor hierarchy}
\label{sec:infinity}

The dga $({\cal T}^\bullet \V^*, \owedge, \rd_L)$ is freely generated. It can thus be viewed as defining a dg- or Q-manifold over a point, or, in other words, it defines a Lie infinity algebra. To describe it, we will use the conventions already employed in section \ref{sec:Lie2}, where the underlying complex of a Lie 2-algebra is concentrated in degrees -1 and -2. If one insists that an ordinary Lie algebra is of degree zero, one needs to shift all of our vector spaces up by one degree, with the disadvantage that then the collection of brackets $l_1,l_2,\ldots$ have a non-homogenous degree; in our case, they all have degree -1. And the degree + 1 derivation operator $\rd_L$ contains all the information about these brackets.

First we need to unravel the underlying vector spaces in the complex $L_\bullet= \oplus_{i \in \mathbb{N}}L_{-i}$ that defines the Lie infinity algebra. Degree one elements in ${\cal T}^\bullet \V^*$ are elements in $\V^*$ and can be viewed as (linear) functions on $\V[1]$, i.e.\ on $\V$ shifted in degree so that its (dual) elements have degree minus one. Thus, $L_{-1}=\V[1]$. Now, the odd shuffle product $\owedge$ coincides with the ordinary wedge product when applied to two elements $\omega_1, \omega_1' \in \V^*$: $$\omega_1 \owedge \omega_1' = \omega_1\wedge \omega_1'\equiv \omega_1 \otimes \omega_1' - \omega_1'\otimes \omega_1.$$ Thus the product of two degree one functions gives a function of degree 2, but not all of them. Evidently, only the subspace $\Lambda^2\V^*\subset {\cal T}^2\V^*$ is covered in this way. To generate all elements of degree two, we need to choose a complement, of which $S^2\V^*$ seems the most natural choice. Thus, $L_{-2}=(S^2\V) [2]$.

To find the vector space at level -3 is already more intricate. For this we need to find a complement to what is generated by the right-hand side of Equation \eqref{shuffle12}. To do so, we make use of the following identity, which one may verify by a straightforward calculation:
\ba e^K \otimes e^M \otimes e^N &=& \frac{1}{4} \left(  [e^K, e^M]_+ \owedge e^N+ [e^M , e^N]_+ \owedge e^K
 \right) +  \frac{1}{6} \left( e^K \owedge e^M \owedge e^N\right) \nonumber \\
{} &+&  \frac{1}{6} \left( [[e^K,e^M]_+,e^N]_- -  [[e^M,e^N]_+,e^K]_-\right) \, , \label{decomposition}
\ea
where we introduced the notation $[e^K,e^M]_\pm = e^K \otimes e^M \pm e^M \otimes e^K$ for the anti-commutator and commutator, respectively. The first line on the right-hand-side of \eqref{decomposition} is seen to be generated by means of lower degree elements and the odd shuffle product. The second line, on the other hand, generated by the
anti-commutator succeeded by the commutator, is thus seen to be a possible complement and can be used as a basis to the dual of $L_{-3}$.

One observes that (a third of) $[[e^K,e^M]_+,e^N]_-$ is the result of the projector $\pi := \frac{1}{3}(\mathrm{id} - t_{13}) (\mathrm{id} + t_{12})$ applied to $e^K \otimes e^M \otimes e^N$, where $t_{ij}$ denotes the permutation of the i-th with the j-th entry. $\pi$ can be identified with the projector to one of the two hook Young tableaux that occur in the decomposition
\be  \yng(1) \otimes \yng(1)\otimes \yng(1) = \yng(3) \oplus \yng(2,1) \oplus \yng(2,1) \oplus \yng(1,1,1)\, , \label{Young}\ee
namely the standard tableau labeled as follows:  {\tiny\young(12,3)}; here we use conventions that the symmetrization corresponding to lines is applied before the antisymmetrization corresponding to columns. Thus, the first three degrees of $L$ can be identified with Young tableaux as follows:
\be  L_{-1} \oplus L_{-2} \oplus L_{-3} \oplus\ldots \cong  \yng(1)[1]\oplus \yng(2)[2] \oplus \yng(2,1)[3]  \oplus \ldots \, \label{symm}\ee
where ${\tiny \yng(1)}$ now corresponds to $\V$ (and no more $\V^*$ as in \eqref{Young} above) and the numbers in the brackets denote the shift in degree.


We mention this relation to Young tableaux, since the fields of the tensor hierarchy were related to them in the literature, at least at lowest levels, see, e.g., \cite{H7}. There is, however, a much more efficient  description of the fields as we are going to explain now, which, moreover, permits an immediate extension to all levels. It works as follows: Let
\be  L_\bullet := FL(\V[1]) = \V[1] \oplus [\V[1],\V[1]]\oplus  [[\V[1],\V[1]],\V[1]] \oplus \ldots  \label{FL}
\ee
be the free super-graded Lie algebra generated by the odd vector space $\V[1]$. In a graded Lie algebra $L_\bullet$, homogeneous elements $a \in L_p$ and $b \in L_q$ satisfy
\be   [a,b] = -(-1)^{pq}[b,a] \, .
\ee
Since elements in $\V[1]$ have degree -1 and are odd, this explains why the beginning of the expansion in \eqref{FL} has symmetry properties and degrees precisely in accordance with \eqref{symm}. Functions on a graded vector space are just polynomials, which in turn can be identified with graded symmetric powers of the dual; in this identification, the point-wise multiplication $\cdot$  of polynomials corresponds to the graded symmetric tensor product of the tensors. In particular, here we have
\be \mathrm{Fun}(L_\bullet) = \mathrm{Sym}(L_\bullet^*)\, . \label{1}
\ee
One now can prove the following isomorphism of graded algebras:
\be ({\cal T}^\bullet \V^*, \owedge ) \cong ( \mathrm{Fun}(L_\bullet), \cdot) \: . \label{2}
\ee
This identification permits also to translate the compatible Loday-differential $\mathrm{d}_L$ to a nilpotent, degree +1 vector field $Q_L$ on the graded manifold $L_\bullet$. For the lowest two levels we will do this explicitly in the subsequent subsection. But further  and more general mathematical details on these facts will be provided in \cite{KSmath}. Here we do not want to overload the present article with mathematical proofs, but instead convey the main underlying mathematical ideas as they find application in the tensor hierarchy and their higher gauge theories in the end.

In particular, the \emph{gauge fields of the tensor hierarchy} can now be viewed collectively as a degree preserving map
\be a^{\infty} \colon (T[1] M,\mathrm{d}) \to (L_\bullet,Q_L) \, .  \label{gaugefield} \ee
Here $M$ is the spacetime and $T[1]M$ its tangent bundle with the additional information that fiber-linear functions on it have degree 1 such that $(\mathrm{Fun}(T[1] M),\cdot) \cong (\Omega^\bullet(M),\wedge)$.
The source of the map $a^{\infty}$ is even canonically a dg- or Q-manifold when adding the nilpotent vector field $Q$, the de Rham differential $\mathrm{d}$, to the data. Thus, the gauge fields are differential forms with values in the graded free Lie algebra $L_\bullet =FL(\V[1])$.

The map \eqref{gaugefield} preserves degrees but not necessarily also the Q-structures; otherwise it would correspond to a generalization of \emph{flat} connections to higher form degrees, which is potentially interesting only for topological theories. Another remark: while the degrees of generators of functions on $L_\bullet$ is unbounded from above, differential forms on $M$ are bounded in their degrees by the dimension $d=\dim M$. Thus, for a fixed choice of $M$, the tower of gauge fields is automatically cut correspondingly. The free Lie algebra $L_\bullet$, on the other hand, is a kind of universal model for them, independent of any choice of dimensions of spacetime $M$. The degrees and commutation properties of its functions are in one to one correspondence with the degrees and commutation properties with respect to the wedge product of the gauge field differential forms. Finally, the fact that $a^{\infty}$ does not preserve the $Q$-structures on the level of the action functional $S$ implies also that the gauge fields do not depend in any way on the chosen Leibniz algebra. The information on the Leibniz algebra only enters the definition of the generalization of the curvature differential forms, the ''field strengths'' of the gauge fields, and in particular the interactions of the gauge fields inside $S[a]$. For some more details on higher gauge theories and their relation to Q-manifolds we refer the reader to \cite{Gruetzmann,Qbundles,Zucchini}.

We again conclude these considerations with some expressions in a basis, i.e.\ in coordinates: Let $e_M$ be a basis of $\V$ dual to  $e ^M$.  Denote the elements in $\V[1]$ corresponding
to $e_M$ by $\xi_M$, and the linear functions corresponding to $e^M$ by $\xi^M$. Then $\deg(\xi_M)=-1$ and $\deg(\xi^M)=1$. We now choose coordinates in the graded manifold $L_\bullet$. According to \eqref{FL}, at degree 1 we can take the odd variables $\xi^M$. At degree 2
we thus get variables $\xi^{MN}=\xi^{NM}$, at degree 3 variables with three indices, $\xi^{MNK}$ with symmetry properties as following from  \eqref{FL} or, equivalently, from the level three Young-tableau in \eqref{symm}. The functions on $L_\bullet$ are generated by these ones,
\be  \mathrm{Fun}(L_\bullet) =\big \langle
\xi^M,\xi^{MN},\xi^{MNK},\ldots \big\rangle \, .\label{generators}
\ee
But, by construction, these are not independent coordinates. To obtain such, we need to choose a basis in the graded free Lie algebra, like a Hall basis consisting of Lyndon words. This is the case since, for example at level two, we have the relation $\xi^{MN}+\xi^{NM}=0$, by the symmetry property of these linear functions. If one wants ''coordinates in the usual sense''---and thus a gauge field for each Lyndon word of the corresponding form degree---one needs to fix some order of the indices; in this case, for example, $M \leq N$. While this has the advantage of constituting a true basis (or a true coordinate system of the graded manifold), it has the disadvantage of making some formulas more clumsy: a term such as $\xi^{MN} \lambda_{MN}$ for some combination of objects carrying two free indices now summarized by $\lambda$,  becomes equal to $\sum_M \xi^{MM} \lambda_{MM}+2\sum_{M<N}\xi^{MN} \lambda_{MN}$, thus with a different prefactor for diagonal terms. We therefore stick to the ''coordinates with symmetry relations''; last but not least, these were  used in the literature on the tensor hierarchy and in fact with a notation receiving a transparent reexplanation in terms of multiple commutators used to generate the free graded Lie algebra of an odd vector space.

\subsection{Truncation to finite $n$ and $n=2$ revisited}\label{sec:truncation}

While the restriction to a certain spacetime dimension $\dim M=d$ automatically cuts the fundamental gauge fields to form degree up to $d$, there are several occasions where one wants to truncate the tower at a lower level already. One of these reasons may be dualities between gauge fields in which case one truncates the tower so as to arrive at  gauge fields of degrees up to $n=[\frac{d}{2}]$, the integer part of half of the dimension of $M$, or even only up to $n=[\frac{d}{2}]-1$. Note, however, that such considerations automatically assume particular properties of the action functional and, moreover, are not global in general if $M$ has a non-trivial topology. Another reason may be much simpler: We consider Yang-Mills gauge theories in all possible dimensions of spacetime. Actually, standard physics on the fundamental level is formulated in this way evidently. Likewise, we may be interested, e.g., in a non-abelian generalization of a gerbe in various dimensions of spacetime, in which case we need to truncate the Lie infinity algebra constructed above to a Lie 2-algebra.

One needs some care to truncate a Lie $m$-algebra to a Lie $n$-algebra, for some $m \in \mathbb{N} \cup \{ \infty\}$ and some $n<m$. For example, take the case $n=1$. Naively truncating the Lie 2-algebra that we found in section \ref{sec:Lie2} to the degree -1 vector space $\V[1]$ would not give a Lie 1-algebra, which should be an ordinary Lie algebra after all (just shifted in degrees here for convenient conventions when extended to higher $n$); indeed, the 2-bracket defined on $\V$ did not satisfy the Jacobi identity and it was precisely this fact that we used  as argument to introduce the space $\W[2]$ as well as  a 1- and 3-bracket.\footnote{That higher gauge theories need higher Lie algebras was observed  e.g.\ in \cite{Baez}. For a general argument leading to this notion in the context of gauge theories, we refer the reader to \cite{Gruetzmann}.} On the other hand, there \emph{is} an ordinary Lie algebra canonically associated to every Leibniz algebra, namely $\gL$. It is the quotient space of $\V$ by the vector space of squares $\S$, which we introduced in the previous section.

This observation generalizes \cite{Lie4}: to truncate a Lie $m$-algebra like $V_\bullet = \sum_{i=1}^m V_{-i}$ to a Lie $n$-algebra for some $n<m$, one removes all the vector spaces $V_p$ with $p<-n$ and considers \be V^{trunc}  = (\sum_{i=1}^{n-1} V_{-i} ) \oplus W\, , \label{Vtrunc} \ee where $W = V_{-n}/\ker (l_1)$ is concentrated in degree -n.  We remark in parenthesis that also $W:=V_{-n}/\mathrm{im} (l_1)\supset V_{-n}/\ker (l_1)$ permits  a consistent and in general larger truncation to a Lie n-algebra; however, the physically relevant one describing a finite tensor hierarchy is the previous one and we restrict our discussion to this choice therefore.

To see that such a truncation is consistent in general, we use again the dual language, representing the Lie infinity algebra as a Q-manifold. Our original Q-manifold is $(L_\bullet, Q_L)$ which is infinite dimensional, although at each degree a finite dimensional vector space. To truncate it to a Q-manifold $({\cal M},Q)$ where the generators of the functions go up to degrees $n$ only, we proceed as follows: Take the generators from degree 1 to $n-1$ in \eqref{generators}. Then apply $Q_L$ to the generators of degree $n-1$. This gives polynomial functions of degree $n$.  Take the part of the polynomial which is of polynomial degree one only. It is necessarily a linear combination of the generators of degree $n$ in \eqref{generators} (and nothing more complicated). Change coordinates at degree $n$ in \eqref{generators} such that the first part of them span this image of $Q_L$ and the second part of them defines a possible complement for a basis. Add the first part of these (new) degree $n$ generators to the generators of ${\cal M}$. Now $Q$ is just the restriction of $Q_L$ to $\mathrm{Fun}({\cal M})$. Evidently this is again a freely generated dga, with generators up to degree $n$, thus giving rise to a Lie $n$-algebra.  Its underlying complex is the one described in the previous paragraph (for $V_\bullet:=L_\bullet$). Geometrically the situation corresponds to a surjective projection \be \pi \colon (L_\bullet, Q_L) \to ({\cal M},Q) \label{Qbundle}
\ee  which respects the Q-structure, $\pi^* \circ Q = Q_L \circ \pi^* $ on $\mathrm{Fun}({\cal M})$; in other words, it defines a (particularly simple) Q-bundle \cite{Qbundles}.

We illustrate this procedure for $n=2$, in which case we will reobtain the Lie 2-algebra introduced in Section \ref{sec:Lie2}. In this way we will implicitly also complete the missing parts in the proof of the validity of the higher Jacobi identities; they are all comprised within the single identity $Q^2=0$, which in turn now follows, by construction,  simply from Loday's old observation
$(\mathrm{d}_L)^2=0$. For this we now need the lowest parts of the nilpotent vector field $Q_L$ in terms of the coordinates \eqref{generators},
to be retrieved from the general formula \eqref{eq:dL} for $\mathrm{d}_L$. We first display the result of the calculation:
\begin{eqnarray} Q_L &=& \left( \frac{1}{2}X_{MN}{}^I \xi^M \xi^N+ X_{MN}{}^I \xi^{MN}\right) \frac{\partial}{\partial \xi^{I}}\nonumber \\
&&+\left( \frac{1}{6}X_{MN}{}^{(I}\xi^{J)}\xi^N\xi^M +  X_{MN}{}^{(I}\xi^{J)N}\xi^M- X_{MN}{}^{(I}\xi^{J)NM}    +\xi^{MN(I}X_{MN}{}^{J)}
\right) \frac{\partial}{\partial \xi^{IJ}}\nonumber \\
&& + \ldots \label{Q_L}
\end{eqnarray}
where the points in the last line correspond to terms containing derivatives with respect to variables of degree at least three.
The first line of equation \eqref{Q_L} follows directly from $\rd_L e^I =  X_{MN}{}^I e^M \otimes e^N \equiv \frac{1}{2}X_{MN}{}^I e^M \owedge e^N + X_{MN}{}^I \frac{1}{2}[ e^M , e^N]_+$. The second line follows in a similar fashion after replacing the basis vectors in the last line of \eqref{dLIJ} by their decomposition \eqref{decomposition}. Here we used a normalization such that $\xi^I$, $\xi^{IJ}$, and $\xi^{IJK}$ correspond to $e^I$, $\frac{1}{2}[ e^I , e^J]_+$, and $\frac{1}{3}[[ e^I , e^J]_+,e^K]_-$, respectively, inheriting their symmetry properties correspondingly. In particular, for example, $\xi^{IJK}=\xi^{JIK}$.

We already made a remark about these generalized type of coordinates at the end of the previous subsection. To extract the brackets on the Lie infinity algebra, we now need appropriately dual coordinates to those in  \eqref{FL}. Since we will restrict our attention to a projection to a Lie 2-algebra only in what follows, we content ourselves with those of degree -1 and -2, $\xi_I\in L_{-1}$ and $\xi_{IJ} \in L_{-2}$. Let us denote the brackets of various degrees on $L_\bullet$ by $l_1$, $l_2$, etc. Then the terms of $Q_L$ displayed in \eqref{Q_L} provide all the information that will be needed for the construction of the truncated Lie 2-algebra. We recall that $l_1$ is the differential in the complex underlying the Lie infinity algebra; here
\be \ldots\stackrel{l_1}{\longrightarrow}  L_{-3} \stackrel{l_1}{\longrightarrow}L_{-2} \stackrel{l_1}{\longrightarrow}L_{-1} \stackrel{l_1}{\longrightarrow}0\, . \label{complexl}\ee
For us it will be sufficient to determine $l_1$ restricted to $L_{-2}$, $l_2$ evaluated on two elements of $L_{-1}$ or on one element of $L_{-2}$ and one of $L_{-1}$, and $l_3$ when evaluated on three elements of $L_{-1}$. This is the sufficient, since all $l_k$ have degree -1 in our case and all other maps or results will be projected to zero in the process of the truncation \eqref{Vtrunc} for $n=2$.

The way of how to extract the brackets $l_k$ on $L_\bullet$ from the vector field $Q_L$, can be summarized as follows: Suppose you have (graded, homogenous, linear, honest) coordinates denoted collectively by $q^\alpha$. Compose the vector field $Q_L$ according to its polynomial degrees, $$Q_L = \sum_{k>1} \frac{1}{k!} C^\beta_{\alpha_1 \ldots \alpha_k} q^{\alpha_1}
\ldots q^{\alpha_k} \frac{\partial}{\partial q^\beta} \, , $$
where the $C$'s are numerical constants defined in this way.
Denote by $q_{\alpha}$ the dual coordinates, which can be viewed also as (homogenous) elements in $L_\bullet$. Then $l_k$ evaluated on the elements $q_{\alpha}$ follow from the simple formula:\footnote{See, e.g., Appendix A in \cite{Gruetzmann} for a proof. But see also \cite{Voronov}.}
$$ l_k(q_{\alpha_1}, \ldots , q_{\alpha_k}) := C^\beta_{\alpha_1 \ldots \alpha_k}  q_{\beta} \, . $$
Applying this to \eqref{Q_L} above, we obtain
\begin{eqnarray} l_1(\xi_{MN}) &=& X_{(MN)}{}^L \xi_L \, , \\
l_2(\xi_{M},\xi_{N}) &=& X_{[MN]}{}^L \xi_L \, ,\\
l_2(\xi_{M},\xi_{NP}) &=& X_{M(P}{}^L \xi_{N)L}\, ,\\ 
l_3(\xi_{M},\xi_{N},\xi_P) &=& -X_{[MN}{}^L \xi_{P]L} \, ,
\end{eqnarray}
where the brackets $( \ldots )$ and $[ \ldots ]$ around indices refer to their (proper) symmetrization and antisymmetrization, respectively. In our context with coordinates of symmetry properties, the coordinates $\xi_{MN}$ are chosen to also be symmetric and to satisfy the following contraction normalization,
\be \langle \xi^{MN}, \xi_{PQ} \rangle = \delta^{(M}_P \delta^{N)}_Q \, . \label{norm} \ee
Since $\xi^{MN}$ corresponds to $\frac{1}{2} e^M \vee e^N \in S^2\V^*$,  \eqref{norm} implies that $\xi_{MN}$ corresponds to $\frac{1}{2} e_M \vee e_N \in S^2\V$.
These identifications now permit us to deduce the coordinate independent meaning of the brackets from the above expressions. On the lowest two levels, the complex \eqref{complexl} can be identified with
\be \ldots\stackrel{l_1}{\longrightarrow} S^2\V \stackrel{l_1}{\longrightarrow} \V \stackrel{l_1}{\longrightarrow}0\, . \label{complexl'}
\ee
Then, for all $u,v,w \in \V$, one has\footnote{General formulas of this sort for arbitrary degrees and multi-brackets will be presented in \cite{KSmath}.}
\begin{eqnarray} l_1(\tfrac{1}{2} u \vee v) &=& u \bullet v \, , \\
l_2(u,v) &=& u \star v  \, , \label{l2uv}\\
l_2(u,v \vee w ) &=&\tfrac{1}{2}\left( (u \circ v) \vee w + v \vee (u \circ w) \right)\, , \label{l2compli}\\ 
l_3(u,v,w) &=& -\tfrac{1}{6} u \vee (v \ast w) + \mathrm{cycl.} \label{l3}
\end{eqnarray}
Here, as already before, $\bullet$ and $\ast$ denote the symmetric and antisymmetric part of the Leibniz product $\circ$. In particular, we see that
 \be l_1\vert_{S^2\V} = s \, , \label{ls} \ee
 where the map $s$ was defined  in Equation \eqref{s}, and that $l_2$ when evaluated between two elements of $\V$ coincides with the antisymmetric part of their Leibniz product.

We are now in the position to describe the truncation \eqref{Vtrunc} for $n=2$ in detail. At degree -1 we keep the vector space $\V$ in \eqref{complexl'}. At level -2, on the other hand, we have to replace $S^2\V$ by its quotient with respect to the kernel of the map \eqref{ls}. (We remark in parenthesis that $S^2 \V/\ker(s)$ is dual to $s^*(\V^*)$, where $s^*$ denotes the map dual to $s$, and (a basis of) this space is precisely what we are supposed to add as generators in degree 2 to the functions of the  Q-manifold $({\cal{M}},Q)$ which corresponds to the Lie 2-algebra). This quotient is in fact isomorphic to the image of $s$ inside $\V$, $\S \equiv s(S^2\V) \subset \V$. Denote by $\W$ a copy of this subspace $\S$, but put in degree -2, and $t \colon \W \to \V$ the corresponding embedding map (which then becomes a map of degree +1); this reproduces indeed the 2-term complex \eqref{VW} of section \ref{sec:Lie2} for the case $\W \cong \S$. For every $x \in \S$, we denote the corresponding vector in $\W$ and of the correspondingly shifted degree by $\widetilde{x}$, as  introduced there already.

Let us now determine the multi-brackets of the truncated Lie infinity algebra and denote them by $(l_k)^{trunc}$. Evidently $(l_1)^{trunc}=t$. Next, $(l_2)^{trunc}(u,v)=l_2(u,v)$, cf.\ Eq.\ \eqref{l2uv}.
To obtain the truncated 3-bracket, we replace the right-hand side of Eq.\ \eqref{l3}, which lives inside $S^2\V$, by its image with respect to $s$ inside $\S\subset \V$. Putting a tilde over the result, with the significance explained above, we obtain
\be \label{l3trunc} (l_3)^{trunc}(u,v,w) =  -\tfrac{1}{3} \widetilde{u \bullet (v\ast w)} + \mathrm{cycl}(u,v,w) \, .
\ee
Applying the analogous procedure for the reduction of \eqref{l2compli}, one obtains after a short calculation using the Leibniz identity \eqref{eq:Leib}
\be \label{l2complitrunc} (l_2)^{trunc}(u,\widetilde{v \bullet w}) = \widetilde{u \bullet (v \bullet w)}\, . \ee
Comparing with the formulas presented in section \ref{sec:Lie2}, we see that the maps $(l_k)^{trunc}$ for $k = 1,2$, and $3$ coincide with $[\cdot]$, $[\cdot,\cdot]$, and $[\cdot,\cdot,\cdot]$, respectively.

In the case  $\W \cong \S$, this also completes the proof that these equations define a Lie 2-algebra. For the bigger one, resulting from the choice $\mathfrak{i} := Z_L(\V)$, cf.\ footnotes \ref{newfootnote} and \ref{footnoteHohm}, one still needs another argument (or perform an explicit verification of the validity of the higher Jacobi identities as it can be found in \cite{ChinesenLie2}). Such an argument can be provided as follows: There is a simple Lie 2-algebra \cite{Hohm} defined on
\be \V \stackrel{\mathrm{id}}{\longrightarrow} \V \label{triv}\ee
for \emph{every} vector space $\V$ equipped with an antisymmetric bracket. The Lie 2-algebra for $\mathfrak{i} := Z_L(\V)$ follows by restriction to the subspace $Z_L(\V)\subset \V$ in the first copy of $\V$ in \eqref{triv}.\footnote{To see this we observe that for $x \in \V$ and $\tilde{y} \in Z_L(\V)$, one has $x \bullet \tilde{y}= x \star \tilde{y}$.} We will now provide a generalization of the Lie 2-algebra \eqref{triv} of \cite{Hohm}, which also simplifies the proof of validity of the structural identities to be satisfied.

Let $\CN$ be a $\Z$-graded manifold and $\xi$ a degree 1 vector field on it, which is not necessarily homological (one could call $(\CN,\xi)$ an almost Q-manifold). Then
\be (T[1]\CN, \rd + L_\xi - \iota_{[\xi,\xi]}) \ee
always is a Q-manifold, as one verifies easily using $L_\xi \equiv [\iota_\xi,\rd]$ and $[L_\xi, L_\xi]= L_{[\xi, \xi]}\equiv [\iota_{[\xi, \xi]},\rd]$.  This shows that every almost $L_\infty$-algebra can be found inside an $L_\infty$-algebra of twice its original size. The result from \cite{Hohm} now follows for the special case that $\CN := \V[1]$ equipped with the vector field $\xi$ corresponding to the initial, anti-symmetric product.

All the Lie 2-algebras discussed in section  \ref{sec:Lie2}, but also \eqref{triv} above, share the property that  $t \colon \W \to \V$ is an embedding: the left center is embedded and all the other ones follow from a restriction of this choice for $\W$ to a smaller sub-vector space (cf.\ also Fig.\ 1). The above Lie infinity algebra, on the other hand, also permits another truncation to a Lie 2-algebra, the underlying complex of which has the form
\be
S^2\V/\mathrm{im}(l_1) \longrightarrow \V \, . \label{bigger}
\ee
If the cohomology of $l_1$ is non-trivial at level two, this gives a bigger Lie 2-algebra than the one with $\W \cong \S$.\footnote{For example, if the Leibniz product $\circ$ vanishes identically, then $l_1 = 0$, there is no quotient to be taken in \eqref{bigger}, and one is left with the 2-term complex $S^2\V \to \V$,  while, on the other side, $\S$ reduces to the zero vector space.}  It can, however, not
coincide with the one where $\W \cong Z_L(\V)$, nor with any of its restrictions, since now the map \eqref{bigger} has a kernel (given precisely by the $l_1$-cohomology).

The Lie 2-algebra appearing in the physical context for the tensor hierarchy truncated at $n=2$ is, however, always the one where $\W \cong \S$. In the following subsection we look at such a context in some more detail.

\subsection{The gauge field sector of gauged maximal supergravity in $d=4$}
\label{sec:new}

We conclude this article with an example from the bosonic gauge field sector of gauged $N=8$ supergravity in four space-time dimensions. By duality arguments, cf., e.g., \cite{Henning}, one restricts to $n=2$ in this case. The Lie group $G$ is the maximally non-compact real form $E_{7(7)}$ of $E_7$ and the subgroup $G_0$ can be, e.g., $G_0=SO(8)$ inside the maximal compact subgroup $SU(8)$ of $E_{7(7)}$. This is only an example for the choice of $G_0$, but we stick to this choice within this subsection; it is, however, a maximal dimensional choice for the image of the embedding tensor, which can be at most 28-dimensional.

The vector space $\V$, in which the 1-form gauge fields $A$ take values in the fundamental representation ${\bf 56}$ of $E_{7(7)}$.  As mentioned in section \ref{sec:theta}, the embedding tensor can be viewed upon as a $G_0$-invariant element in $\V^* \otimes \g_0 \subset \V^* \otimes \g$. It is only here where the representation theory of $\g$ plays a role now. $G$ and thus also the adjoint representation $\g$ has dimension 133 in this case. Decomposing ${\bf 56} \otimes {\bf 133}$ into its irreducible $\g$-representations, one has: ${\bf 56} \oplus {\bf 912} \oplus {\bf 6480}$. Supersymmetry requires the tensor $\theta$ to lie inside ${\bf 912}$, which is also compatible with the above mentioned choice for $\g_0$. There is still some freedom in choosing the $\g_0$-equivariant $\theta \colon \V \to \g_0 \subset \g$. Any choice fixes the Leibniz algebra $(\V\cong \R^{56},\circ)$, where the Leibniz product is defined by means of equation \eqref{eq:reconstr}. We remark in parenthesis that in this language, the representation theoretic constraint on $\theta$ above can be also reexpressed \cite{local}  in a purely algebraic  form \be \Omega(x \circ x, x)=0 \qquad \forall x \in \V \, , \label{Omega} \ee
where $\Omega$ is a $G$-invariant symplectic form on $\V$, which corresponds to the well-known embedding $E_{7(7)} \subset \mathrm{Sp}(56)$.  $\Omega$ is thus in particular also $G_0$-invariant, which can be equivalently be expressed as saying that it is invariant with respect to the left action of $\V$ onto itself, \be \Omega(x \circ y, z) + \Omega(y, x \circ z)=0 \label{Omegainv}\ee  for all $x,y,z \in \V$.


In the abstract tensor hierarchy, cf.\ equations \eqref{gaugefield} and \eqref{symm}, one has a 1-form gauge field $A^{\infty}$ with values in $L_{-1}[-1]\cong \V$, a 2-form gauge field $B^{\infty}$ with values in $L_{-2}[-2]\cong S^2\V$, a 3-form field $C^{\infty}$ with values in $L_{-3}[-3]$ etc. According to what we said above, we need to consistently truncate this infinite tower of gauge fields at the level $n=2$ to arrive at 2-level truncated tensor hierarchy.

The relevant Lie 2-algebra is then defined on
\be \S \hookrightarrow \V \, , \ee
where from now on we regard the vector space on the left as a sub-vector space of the one on the right, sparing us in what follows to depict the map $t$ as in \eqref{VW}.
The field content of the gauge sector of the SUGRA under discussion is then provided by the following truncated tensor hierarchy, consisting only of the following two levels:
 \be A = A^{\infty} \in \Omega^1(M,\V) \qquad \mathrm{and} \qquad B = s(B^{\infty}) \in  \Omega^2(M,\S) \, .
 \label{fields}\ee
 In the supergeometric language, this means that the gauge fields of the $(n=2)$-truncated theory correspond to a morphism of graded manifolds
 \be a^{trunc} \colon T[1] M \to \CM:= \V[1] \times \S[2] \ee
 replacing the previous \eqref{gaugefield}. In fact, $a^{trunc} = \pi \circ a^{\infty}$, where $\pi$ is the corresponding projection map \eqref{Qbundle}.
 The truncation is to $\S$ and not, e.g., to the bigger Lie 2-algebra \eqref{bigger}, since one observes that in the action functional of the four-dimensional gauged maximal supergravity theories, the 2-form fields $B^\infty$ always appear under the image of the map $s$, i.e.,  for what concerns the gauge fields, they are precisely formulated in terms of the fields \eqref{fields}.

Let us now determine some bounds on the dimension of the vector space $\S$. First, we know that $\dim \V = 56$ and $\dim \g_0 = \dim so(8) = 28$. This already implies that
\be 0 \leq k:=\dim\S \leq 28 \, , \label{bound}
\ee
since we know that the kernel $\W$ of $\theta \colon \V \to \g_0$ contains the vector space $\S$, cf.\ Fig.\ 1. We remark as an aside that $\g_0$ is also the minimal Lie algebra $\g_{\mathrm{min}}$ of the corresponding Leibniz algebra, i.e.\ the maximal quotient of $\g_{\mathrm{Lie}}$; this is the case, since the fundamental $\g$-representation $\V$ remains effective (faithful) also upon restriction to $\g_0$. Which dimension $k$ the vector space $\S$ has, depends on the choice of the embedding tensor or Leibniz algebra. In general, we only know about the bound \eqref{bound} and that, correspondingly, $\dim(\g_{\mathrm{Lie}})=56-k$.

 In the literature one finds, cf., e.g., \cite{Henning}, that the field content consists of
$A^{lit} \in \Omega^1(M,{\bf 56})$ and $B^{lit}\in \Omega^2(M,{\bf 133})$, where ${\bf 133}$ denotes the adjoint representation of $\g$. The argument given is that, on the one hand, $B^\infty\in \Omega^2(M,S^2\V)$ and that, on the other hand, the constraint \eqref{Omega} implies that inside
\be {\bf 56}\vee {\bf 56}= {\bf 1463}\oplus {\bf 133}\ee
only the second summand is selected when projecting by means of the embedding tensor $\theta$. Let us look at this in more detail: First we observe that the Leibniz product \eqref{eq:reconstr} is $G_0$-equivariant, since $(g \cdot x) \circ (g \cdot  y) = \left(g \cdot \theta(x) \cdot g^{-1} \right) \cdot \left(g \cdot y\right) = g \cdot (x \circ y)$ holds true for all $x,y \in \V$, $g\in G_0$. Thus also its symmetrization, the map $s\colon S^2 \V \to \V$, is. What the above special role of ${\bf 133}$ means is merely that by the choice of $\theta$---or, \emph{equivalenty}, by the constraint \eqref{Omega} satisfied by the Leibniz product on ${\bf 56}$---automatically
\be {\bf 1463} \subset \ker(s) \, . \ee
But not all of the remaining 133 independent components of the 2-form fields enter actively into the gauge sector of this theory: There are only $k$ of them that survive in the end by taking the full quotient of $S^2\V$ with respect to $\ker(s)$, $B=s(B^{lit}) = s(B^{\infty})$, where $k$ is subject to the bounds \eqref{bound}. 

We finally display the relevant part of the action functional for this sector of the theory. For this purpose it is convenient to define the following generalization of the curvatures or field strengths relevant for the $n = 2$ Leibniz theory \cite{gerbe,LeibYM} (but cf.\ also \cite{Olaf-Henning,Olaf-Henning2}):
\ba F &=& \rd A + \tfrac{1}{2}A \circ A - B \,,\label{F}\\
G& =& \rd B - A  \circ F + A \star \rd A\, .\label{G}
\ea
Here we make use of the notation introduced in \eqref{s}, \eqref{a} and wedge products between differential forms are understood, so that, e.g., $A \circ A \equiv A^a \wedge A^b \otimes e_a \circ e_b = A \star A$. The above definitions imply in particular  $F \in \Omega^2(M,\V)$ and
$ G\in \Omega^3(M,\S)$.

The infinitesimal gauge transformations are of the following form:
\ba \delta A &=& \rd \epsilon + A \star \epsilon + \mu \, ,
\label{symm1} \\ 
\delta B &=& \rd \mu + A \bullet \left( \mu -\tfrac{1}{2} A \circ \epsilon\right)+ (F-B)\bullet \epsilon \, ,\label{symm2}
\ea
where $\epsilon \in C^\infty(M,\V)$ and $\mu \in \Omega^1(M,\S)$ are the arbitrary parameters. The field strengths \eqref{F} and \eqref{G}
transform covariantly with respect to them,
\be \delta F = -\epsilon\circ F \quad , \qquad    \delta G = - \epsilon\circ G\, , 
\label{deltaF}\ee
as shown in detail in \cite{LeibYM}.

The relevant part of the action functional then is, at least schematically, of the following form, cf., e.g., \cite{Henning,local,Olaf-Henning2}:
\be \label{S} S[A,B,\ldots] :=  \int_M  \kappa_1(F,\ast F) + \kappa_2(F, F) + \int_{M_5} \kappa_3(F,G) \, ,
\ee
where now some further explanations are in order. First, $(\kappa_i)_{i=1}^3$ are symmetric bilinear forms on $\V$. In fact, for $i=3$ we could have limited ourselves to a bilinear form on $\V \otimes \S$. But it turns out that upon restriction to $\S \otimes \S$ it has to become symmetric, so we can extend it as required without loss of generality. $\kappa_3$ is required to satisfy the following invariance condition:
\be \kappa_3(x \circ y,y)=0 \label{kappa3inv}\ee  for all $x,y \in \V$.  By means of this condition,
the contribution $\kappa_3(F,A \circ F)$ to  $\kappa_3(F,G)$, which contains all the terms without derivatives, vanishes evidently. By an explicit calculation one then shows, that all the remaining terms  in $\kappa_3(F,G)$ assemble into an exact contribution. This is needed so that the last term in \eqref{S}, which is written in Wess-Zumino form with the usual understanding, indeed induces a local term on $M = \partial M_5$.

Second, the dots in \eqref{S} indicate that this part of the functional depends on additional fields. This is, on the one hand, the dynamical metric of the supergravity theory, which is used to define the Hodge duality operation $\ast$ on $M$, entering the first, Yang-Mills type contribution to $S$. But, on the other hand, the bilinear forms $\kappa_i$ are permitted to depend on the scalar fields of the theory. These in turn also transform with respect to the gauge symmetries \eqref{symm1} and  \eqref{symm2}. For $i=1,2$, the correspondingly induced transformation on the bilinear forms have to lead to
\be \left(\delta_{(\epsilon,\mu)} \kappa_i \right)(x,x)= 2\kappa_i(\epsilon \circ x,x) \label{kappa12}
\ee
for every $x\in \V$ and all parameters $\epsilon$ and $\mu$ as specified above, while $\kappa_3$ has to be invariant, $\delta_{(\epsilon,\mu)} \kappa_3=0$. (We remark in parenthesis that, together with \eqref{kappa3inv}, this implies that \eqref{kappa12} holds true also for $i=3$).

With these conditions, the gauge invariance of \eqref{S} is manifest---which was one of the reasons for not spelling out the Chern-Simons-like contributions to it coming from the five-dimensional integral in terms of a four-dimensional representative.

In the supergravitational context, the bilinear forms $\kappa_i$ are typically degenerate, and different choices of them can still lead to locally on-shell equivalent systems. In particular, generically not all of the 56 $A$-fields finally appear in such a concretely chosen functional, and this also holds true for the $B$-fields: once that the Leibniz algebra is fixed, there can be up to $k$ of them, where $k$ is subject to the constraint \eqref{bound}, but generically not all of them will appear inside of $S$ after the choice of the bilinear forms.

Leibniz gauge theories can be considered also independently as particular higher gauge theories in their own right, cf., e.g., \cite{Olaf-Henning,Olaf-Henning2,LeibYM,Olafneu}. In that case, a functional of the form \eqref{S}, to which one may add also
\be  \int_M\kappa_4(G,\ast G) \, ,
\ee
depends only on the fields $A$ and $B$. The Hodge duality is then taken with respect to some background metric on $M$, and gauge invariance requires that all the four, now field-independent bilinear forms, $i=1,\ldots,4$, satisfy
\be \kappa_i(x \circ y,y) =0\ee
for all $x,y \in \V$. 






\acknowledgments
T.S.~is particularly grateful to Henning Samtleben, who spent time and effort in explaining the essential parts of the embedding tensor to him  and who participated also actively in the beginning phase of this work. We furthermore gratefully acknowledge discussions with him on subsection \ref{sec:new}, the addition of which was motivated by questions of one of the anonymous referees.

The research of A.K.~was supported by the grant no.~18-00496S of the Czech Science Foundation and this work in general  by the LABEX MILYON (ANR-10-LABX-0070) of Université de Lyon, within the program "Investissements d'Avenir" (ANR-11-IDEX-0007) operated by the French National Research Agency (ANR).

\appendix\section{String-proof of associativity of the odd shuffle product $\owedge$} \label{sec:app}

Let $V$ be a vector space. Given an element $\phi\in \mathcal{T}^\bullet (V^*)$, we associate to it a functional on the path space $P(V)$, the space of maps from $[0,1]$ to $V$, as follows:
\ba\label{function_on_path_space}
 F_{\phi} [\gamma]=\sum\limits_{n=0}^\infty\int\limits_{\Delta^n} \md^n t \,\langle\gamma (t_1)\otimes \ldots \otimes \gamma (t_n),  \phi\rangle
\ea
for any $\gamma\in P(V)$, where $\langle , \rangle$ is the canonical pairing between $\mathcal{T}^\bullet (V)$ and $\mathcal{T}^\bullet (V^*)$ and
\ba\nonumber
\Delta^n =\{(t_1,\ldots, t_n)\mid 0\le t_n\le \ldots \le t_1\le 1\}\,.
\ea
In other words, $F_{\phi} [\gamma]=\langle \mathcal{P}\exp (\gamma)(1,0),  \phi\rangle$, where $\mathcal{P}\exp (\gamma)(t, t')$ is the path-ordered exponential of $\gamma$ with values in $\widehat{\mathcal{T}^\bullet(V)}$, the formal completion of $\mathcal{T}^\bullet(V)$:
\ba\nonumber
\mathcal{P}\exp (\gamma)(t,t') = \sum\limits_{n=0}^\infty \int_{t'\le t_n\le\ldots \le t_1\le t} \md^n t \,\gamma (t_1)\otimes \ldots \otimes \gamma (t_n) \,.
\ea
Then for any $\phi, \phi'\in \mathcal{T}^\bullet (V^*)$, one has
\ba\nonumber
 F_{\phi\shuffle \phi'}=F_\phi F_{\phi'}\,,
\ea
where $\shuffle$ denotes the (ungraded) shuffle product.
Indeed, let us assume that $\phi\in \mathcal{T}^n (V^*)$ and $\phi'\in \mathcal{T}^m (V^*)$. For any $\gamma\in P(V)$
\ba\label{product_of_functionals1}
\left( F_\phi F_{\phi'}\right) [\gamma]=
\int\limits_{\Delta^n\times\Delta^m} \md^{n+m} t \,\langle\gamma (t_1)\otimes \ldots \otimes \gamma (t_{n+m}),  \phi\otimes\phi'\rangle \,,
\ea
where
\ba\nonumber
\Delta^n\times\Delta^m =\{(t_1,\ldots, t_{n+m})\mid 0\le t_n\le \ldots \le t_1\le 1\,, 0\le t_{n+m}\le \ldots \le t_{n+1}\le 1 \}\,.
\ea
The expression (\ref{product_of_functionals1}) can be rewritten as follows:
\ba\label{product_of_functionals2}
\left( F_\phi F_{\phi'}\right) [\gamma]=
\sum\limits_{\sigma\in Sh_{n,m}}\,\int\limits_{\Delta^{n+m}_\sigma} \md^{n+m} t \,\langle\gamma (t_1)\otimes \ldots \otimes \gamma (t_{n+m}),  \phi\otimes\phi'\rangle \,,
\ea
where
\ba\nonumber
\Delta^{n+m}_\sigma =\{(t_1,\ldots, t_{n+m})\mid 0\le t_{\sigma(n+m)}\le \ldots \le t_{\sigma(1)}\le 1 \}\,.
\ea
Therefore indeed
\ba\label{product_of_functionals3}
\left( F_\phi F_{\phi'}\right) [\gamma]=
\int\limits_{\Delta^{n+m}} \md^{n+m} t \,\langle\gamma (t_1)\otimes \ldots \otimes \gamma (t_{n+m}),  \phi\shuffle\phi'\rangle =F_{\phi\shuffle \phi'} [\gamma]\,.
\ea
Now we show that the morphism of the algebra  $\left(\mathcal{T}^n (V^*), \shuffle\right)$ to the algebra of functionals on $P(V)$ is injective. We can easily verify by rescaling that, if $F_\phi [\gamma]=0$ for all $\gamma\in P(V)$, then $\langle \mathcal{P}\exp (\gamma)(t,0), \phi\rangle =0$ for all $\gamma\colon [0,t]\to V$.
To simplify our life, let us allow piecewise continuous maps to $V$. We can take
\ba\label{piecewise_continuous}
\gamma_{\tau_1, \ldots, \tau_n} (t')=\left\{
 \begin{array}{cc}
 v_1\,, & 0\le t' <\tau_1 \\
 v_2\,, & \tau_1\le t' < \tau_1+\tau_2\\
 \vdots & \\
 v_n\,, &\quad \tau_1+\ldots +\tau_{n-1} \le t' \le t
 \end{array}
\right.
\ea
where $t=\tau_1+\ldots + \tau_n$ and $v_1, \ldots, v_n\in V$ are arbitrary (constant) vectors. We first observe that, if $\gamma \equiv v$ is a constant map, then
\ba\nonumber
 \mathcal{P}\exp (\gamma)(t)= \exp (tv)=\sum\limits_{n=0}^\infty \underbrace{v\otimes \ldots \otimes v}_{n \, \mathrm{times}} \frac{t^n}{n!}\,.
\ea
It follows from the general properties of the path-ordered exponential that $\mathcal{P}\exp (\gamma)(t,t')=\mathcal{P}\exp (\gamma)(t,\tau) \otimes \mathcal{P}\exp (\gamma)(\tau,t')$ for any $\tau$ in the domain of $\gamma$. Therefore for $\gamma$ as in (\ref{piecewise_continuous}) we obtain
\ba\nonumber
 \mathcal{P}\exp (\gamma)(t,0)=\exp (\tau_n v_n)\otimes \ldots \otimes \exp (\tau_1 v_1)\,.
\ea
Let $\phi\in\mathcal{T}^n (V^*)$ be in the kernel of the morphism $F$ to the algebra of functionals on $P(V)$, that is, $F_\phi [\gamma]=0$ for all $\gamma$. Then, in particular, one has
$\langle\exp (\tau_n v_n)\otimes \ldots \otimes \exp (\tau_1 v_1), \phi\rangle=0$ for all $\tau_1, \ldots, \tau_n$ and $v_1, \ldots, v_n\in V$, which implies $\langle v_n\otimes\ldots \otimes v_1, \phi\rangle \equiv 0$ (one should take the $\tau_1\ldots\tau_n -$term in the corresponding power series expansion) and thus $\phi=0$.
As a corollary, the algebra  $\left(\mathcal{T}^n (V^*), \shuffle\right)$ is commutative and associative.

\vskip 2mm\noindent Now we would like to adapt the above statement to the graded super case. Let $L$ be a graded vector space and $\shuffle$ be the graded super shuffle product in $\mathcal{T}^n (L^*)$; we will show that $\left(\mathcal{T}^n (L^*), \shuffle\right)$ is a super commutative associative algebra. Consider a free left graded module $L_{\Lambda}$, generated by $L$ over the Grassmann algebra $\Lambda=\Lambda^\bullet (\xi_1, \xi_2, \ldots)$. The right free graded module $L^*_\Lambda$ is naturally isomorphic to the space of graded $\Lambda-$linear maps $L_{\Lambda}\to\Lambda$, so that $\langle av, \phi b\rangle = a\langle v,\phi\rangle b$
for all $v\in L$, $\phi\in L^*$, $a,b\in\Lambda$. Similarly, $\mathcal{T}^p_\Lambda (L^*_\Lambda)\simeq \mathcal{T}^p (L^*)_\Lambda$ is naturally dual to $\mathcal{T}^p_\Lambda (L_\Lambda)$ over $\Lambda$. Moreover, one has
\ba \nonumber
\langle\sigma (v_1\otimes\ldots\otimes v_n), \sigma (\phi_1\otimes\ldots\otimes \phi_n)\rangle = \langle v_1\otimes\ldots\otimes v_n,  \phi_1\otimes\ldots\otimes \phi_n\rangle
 \ea
 for any $\sigma$ from the group of permutations $\mathbb{S}_n$; here the representation of $\mathbb{S}_n$ in the tensor product $\mathcal{T}^p (L)$ is generated by the action of the elementary permutations $\sigma_{12}$ as follows:
 \ba\label{sign_permutation}\sigma_{12}(v_1\otimes v_2)=(-1)^{\deg (v_1)\deg (v_2)}v_2\otimes v_1\,.
 \ea
 This property is also true for the corresponding free modules over $\Lambda$.

\vskip 3mm\noindent Let $\gamma$ be a path taking values in the even part of $L_{\Lambda}$; it can be also regarded as a morphism of super-manifolds $[0,1]\times\mathbb{R}^{\scriptscriptstyle 0\mid \infty}\to L$. Given $\phi\in\mathcal{T}^\bullet (L)$, we obtain a map on the space of those morphisms with values in $\Lambda$ by the same formula (\ref{function_on_path_space}) as above. Apparently, $\deg (F_\phi)=\deg (\phi)$, which implies in particular $F_\phi F_\psi =(-1)^{\deg (\phi)\deg (\psi)}F_\psi F_\phi$ for any $\phi$ and $\psi$ of pure degree. Now, applying the same arguments as in the even case, see (\ref{product_of_functionals1}), (\ref{product_of_functionals2}), (\ref{product_of_functionals3}),  we immediately verify that $F_{\phi\shuffle \psi}=F_\phi F_{\psi}$. Let us remark that the Koszul sign in the shuffle product for $\mathcal{T}^n (L^*)$ follows from the sign rule (\ref{sign_permutation}).

In the main text, $L = \Pi \V^*$ and  the odd shuffle product $\shuffle$ is denoted by $\owedge$ when applied between elements in  $\mathcal{T}^\bullet (\V^*)$.


\begin{thebibliography}{99}
\bibitem{H1}
 \textsc{H.~Nicolai} and  \textsc{H.~Samtleben},
\emph{Maximal gauged supergravity in three-dimensions},
Phys. Rev. Lett. \textbf{86}, 1686 (2001).
\bibitem{H2}
 \textsc{B.~de Wit},  \textsc{H.~Samtleben} and  \textsc{H.~Trigiante},
\emph{On Lagrangians and gaugings of maximal supergravities},
Nucl. Phys. \textbf{B 655}, 93 (2003).
\bibitem{H3}
\textsc{B.~de~Wit}, \textsc{H.~Samtleben} and \textsc{M.~Trigiante}, \emph{The
  maximal ${D} = 5$ supergravities}, Nucl. Phys., \textbf{B 716}, (2005)
  215--247.
\bibitem{H4}
 \textsc{H.~Samtleben} and  \textsc{M.~Weidner},
\emph{The Maximal D=7 supergravities},
Nucl. Phys. \textbf{B 725} (2005) 383.
\bibitem{H5}
\textsc{B.~de~Wit} and \textsc{H.~Samtleben}, \emph{Gauged maximal
  supergravities and hierarchies of nonabelian vector-tensor systems},
  Fortschr. Phys., \textbf{53}, (2005) 442--449.
\bibitem{H6}
\textsc{B.~de~Wit}, \textsc{H.~Nicolai} and \textsc{H.~Samtleben},
  \emph{Gauged supergravities, tensor hierarchies, and {M}-theory}, JHEP,
  \textbf{0802}, (2008) 044.
 \bibitem{H7} \textsc{B.~de~Wit} and \textsc{H.~Samtleben}, \emph{The end of the $p$-form hierarchy}, JHEP, \textbf{08}, (2008)
  015.
\bibitem{H8}
 \textsc{G.~Dall'Agata} and  \textsc{G.~Inverso},
\emph{On the Vacua of N = 8 Gauged Supergravity in 4 Dimensions},
Nucl. Phys. \textbf{B 859}, 70 (2012).
\bibitem{H9}
 \textsc{G.~Dibitetto},  \textsc{A.~Guarino} and  \textsc{D.~Roest},
\emph{Charting the landscape of N=4 flux compactifications},
JHEP \textbf{1103}, 137 (2011).
\bibitem{Lada1}\textsc{T.~Lada} and \textsc{M.\ Markl},
\emph{Strongly homotopy Lie algebras}.
Communications in Algebra
{\bf 23}, Issue 6 (1995) 2147.
\bibitem{Lada2}\textsc{T.~Lada} and \textsc{J.\ Stasheff},
\emph{Introduction to SH Lie algebras for physicists}.
Int.\ J.\ Theor.\ Phys.\ \textbf{32} (1993) 1087.
\bibitem{Severa} \textsc{P.~\v Severa},
\emph{Some title containing the words "homotopy" and "symplectic", e.g. this one}. Travaux Mathématiques, \textbf{16} (2005), 121.
\bibitem{Gruetzmann} \textsc{M.\ Gr\"utzmann} and \textsc{T.\ Strobl},
\emph{General Yang-Mills type gauge theories for p-form gauge fields:
from physics-based ideas to a mathematical framework or from Bianchi
identities to twisted Courant algebroids.}
Int.\ J.\ Geom.\ Methods Mod.\ Phys.\
\textbf{12} No.\ 01 (2015) 1550009.
\bibitem{Qbundles}  \textsc{A.\ Kotov} and \textsc{T.\ Strobl}, \emph{Characteristic classes associated to Q-bundles}.
      International Journal of Geometric Methods in Modern Physics {\bf 12}  (2015) 1550006.
\bibitem{Lie4} \textsc{S.\ Lavau}, \textsc{H.\ Samtleben} and  \textsc{T.\ Strobl},
\emph{Hidden Q-structure and Lie 3-algebra for non-abelian superconformal models in six dimensions}.
  Journal of Geometry and Physics {\bf 86} (2014) 497.
\bibitem{KSmath} \textsc{A.\ Kotov} and \textsc{T.\ Strobl}, in preparation.
\bibitem{Fiorenza} \textsc{D.\ Fiorenza} and \textsc{M.\ Manetti}, \emph{$L_\infty$ structures on mapping cones}. Algebra Number Theory {\bf 1}  (2007) 301.
\bibitem{Getzler} \textsc{E.\ Getzler}, \emph{Higher derived brackets}. ArXiv:1010.5859.
\bibitem{Palmkvist} \textsc{J.\ Palmkvist}, \emph{The tensor hierarchy algebra}, 	J.\ Math.\ Phys.\ {\bf 55} (2014) 011701.
\bibitem{gerbe}\textsc{T.\ Strobl},
\emph{Non-abelian Gerbes and Enhanced Leibniz Algebras.}
Phys.\ Rev.\ D, Rapid Communications {\bf 94} (2016), no.\ 2, 021702.
\bibitem{Wagemann}\textsc{T.\ Strobl} and \textsc{F.\ Wagemann},
\emph{Enhanced Leibniz Algebras: Structure Theorem and Induced Lie 2-Algebra}, in preparation.
\bibitem{LeibYM}  \textsc{T.\ Strobl}, \emph{Leibniz-Yang-Mills Gauge Theories and the 2-Higgs Mechanism}. ArXiv:1903.07365. To be published in Phys.\ Rev.\ D.
 \bibitem{Luxtalk}  \textsc{T.\ Strobl}, \emph{Mathematics around Lie 2-algebroids and the tensor hierarchy in gauged supergravity}. Talk at "Higher Lie theory", University of Luxembourg, 2013.
\bibitem{Lavau}\textsc{S.\ Lavau}, \emph{Tensor hierarchies and Lie n-extensions of Leibniz algebras}.  ArXiv:1708.07068, to be published in Journal of Geometry and Physics.
 \bibitem{Olaf-Henning}\textsc{O.\ Hohm} and \textsc{H.\ Samtleben}, \emph{Leibniz-Chern-Simons Theory and Phases of Exceptional Field Theory}.  ArXiv:hep-th/1805.03220, to be published in Communications of Mathematical Physics.
 \bibitem{ChinesenLie2}\textsc{Z.\ Liu} and \textsc{Y.\ Sheng},
\emph{From Leibniz algebras to Lie 2-algebras.}
Algebr.\ Represent.\ Theory {\bf 19} (2016) 1.
\bibitem{Hohm} \textsc{O.\ Hohm}, \textsc{V.\ Kupriyanov}, \textsc{D.\ L\"ust} and \textsc{M.\ Traube}, \emph{Constructions of $L_\infty$  algebras and their field theory realizations}. Advances in Mathematical Physics, Article ID 9282905 (2018).
\bibitem{Loday1} \textsc{J-L.\ Loday},
\emph{Une version non commutative des alg\`ebres de Lie: les alg\`ebres de
Leibniz.}
Enseign.\ Math.\ (2) {\bf 39} (1993), no.\ 3-4, 269.
\bibitem{Loday2} \textsc{J-L.\ Loday} and  \textsc{T.~Pirashvili},
\emph{Universal enveloping algebra of Leibniz algebras and (co)homology}.
Math. Ann. {\bf 296} (1993) 139.
\bibitem{Poncin}\textsc{J.\ Grabowski}, \textsc{D.\ Khudaverdyan} and
 \textsc{N.\ Poncin}, \emph{The supergeometry of Loday algebroids}.  Journal of Geometric Mechanics (2013), 5(2), 185.
\bibitem{Zucchini}  \textsc{R.~Zucchini}, \emph{Algebraic formulation of higher gauge theory}.
 Journal of Mathematical Physics, \textbf{58}   (2017) 1.
 \bibitem{Baez} \textsc{J.\ Baez}, \emph{Higher Yang-Mills Theory}. ArXiv:hep-th/0206130.
\bibitem{Voronov}\textsc{T.\ Voronov},
\emph{Higher derived brackets and homotopy algebras}.
Journal of Pure and Applied Algebra
{\bf 202} 1-3 (2005)  133.
\bibitem{Henning} \textsc{H.~Samtleben}, \emph{Lectures on Gauged Supergravity and Flux Compactifications}. Class.\ Quant.\ Grav.\ {\bf 25} (2008) 214002.
\bibitem{local}\textsc{B.~de~Wit}, \textsc{H.~Samtleben} and \textsc{M.~Trigiante}, \emph{Magnetic charges in local field theory}. JHEP {\bf 09} (2005) 016.
\bibitem{Olaf-Henning2}\textsc{O.\ Hohm} and \textsc{H.\ Samtleben}, \emph{Higher Gauge Structures in Double and Exceptional Field Theory}. ArXiv:1903.02821.
\bibitem{Olafneu} \textsc{R.\ Bonezzi} and \textsc{O.\ Hohm}, \emph{Leibniz Gauge Theories and Infinity Structures}. ArXiv:1904.11036.
\end{thebibliography}
\end{document}